\documentclass[preprint,floatfix] {revtex4} 
\newcommand{\rvec}{\mathrm {\mathbf {r}}} 
\newcommand{\pvec}{\mathrm {\mathbf {p}}} 
\usepackage{graphicx}
\usepackage{subfigure}
\usepackage{xcolor}
\usepackage{amsmath}
\usepackage{enumerate}
\usepackage{longtable}
\usepackage{tabularx}

\usepackage{color, soul}
\definecolor{darkblue}{rgb}{0,0,0.5}
\setulcolor{darkblue}

\begin{document}
\title{Information-entropic measures for non-zero $l$ states of confined hydrogen-like ions}

\author{Neetik Mukherjee}
\altaffiliation{Email: neetik.mukherjee@iiserkol.ac.in.}

\author{Amlan K.~Roy}
\altaffiliation{Corresponding author. Email: akroy@iiserkol.ac.in, akroy6k@gmail.com.}
\affiliation{Department of Chemical Sciences\\
Indian Institute of Science Education and Research (IISER) Kolkata, 
Mohanpur-741246, Nadia, WB, India}

\begin{abstract}
%%1234567890 %%1234567890 %%1234567890 %%1234567890 %%1234567890 %%1234567890 %%1234567890 %%1234567890 %%1234567890 %%1234567890
%%Shannon entropy ($S$) in both $r$ and $p$ space for a spherically \emph{confined} hydrogen atom (CHA) has been pursued several times 
%%in literature. However, R{\'e}nyi entropy ($R$), Tsallis entropy ($T$), and Onicescu energy in composite space have been investigated 
%%for some low-lying $l=0$ states of CHA. 
 R\'enyi entropy ($R$), Tsallis entropy ($T$), Shannon entropy ($S$), and Onicescu energy ($E$) are studied in a spherically 
confined H atom (CHA), in conjugate space, with special emphasis on non-zero $l$ states. {\color{red} This work is a continuation of our recently
published work \cite{mukherjee17}.} Representative calculations are done by employing \emph{exact} analytical wave functions in $r$ space. 
Accurate $p$ space-wave functions are generated numerically by performing Fourier transform on respective $r$-space counterparts. 
Further, these are extended for H-isoelectronic series by applying the scaling relations. $R, T$ are evaluated by choosing the order of 
entropic moments $(\alpha, \beta)$ as $(\frac{3}{5}, 3)$ in $r$ and $p$ spaces. Detailed, systematic results of all these measures with 
respect to variations of confinement radius $r_c$ are offered here for arbitrary $n,l$ quantum numbers. For a given $n$, at small 
$r_{c}$, $ R_{\rvec}^{\alpha}, T_{\rvec}^{\alpha}, S_{\rvec}$ collapse with rise of $l$, attain a minimum, then again grow up. Growth in 
$r_c$ shifts the point of inflection towards higher $l$ values. An increase in $Z$ enhances localization of a particular state. 
Several other new interesting inferences are uncovered. Comparison with literature results (available only for $S$ in $2p$, $3d$ 
states), offers excellent agreement.

\vspace{5mm}
{\bf PACS:} 03.65-w, 03.65Ca, 03.65Ta, 03.65.Ge, 03.67-a.

\vspace{5mm}
{\bf Keywords:} R\'enyi entropy, Shannon entropy, Onicescu energy, Tsallis entropy, Confined hydrogen atom, Scaling relation. 
 
\end{abstract}
\maketitle

\section{introduction}
Confinement of an atom or molecule inside an impenetrable cavity was first studied in the fourth decade of twentieth century 
\cite{chen57}. Progress of research on such quantum systems was reviewed several times \cite{chen57,jaskolski96, sabin2009, 
katriel12} recording their importance in both fundamental physics and chemistry as well as in various engineering branches. 
They have relevance in many different physical situations, e.g., atoms under plasma environment, impurities in crystal lattice 
and semiconductor materials, trapping of atoms/molecules in zeolite cages or inside an endohedral cobweb of fullerenes, 
quantum wells, quantum wires, quantum dots \cite{sen2014electronic} and so forth. Furthermore, such models were designed to 
mimic the high pressure environment inside the core of planets. Also, they have contemporary significance in interpreting 
various astrophysical phenomena \cite{pang11} and many other interesting areas. 
  
Theoretical study of a Hydrogen atom within an infinite spherical cavity was first published in 1937 
\cite{michels37}.  Over the years, this simple confined hydrogen atom (CHA) model has served as a precursor to improve our 
understanding about the consequences of confinement in atomic electronic structure. In last decade, a CHA under the influence 
of various restricted environment has been extensively followed. Majority of these investigations include trapping of H atom 
either in a spherical box of penetrable, impenetrable walls or inside a hard box of different geometrical shape and size 
\cite{katriel12,aquino13,jiao17,coll17,centeno17}. In the realm of atomic physics, CHA provides us with many attractive physical and chemical 
properties. Numerous theoretical methods like perturbation theory, Pad\'{e} approximation, WKB method, Hypervirial theorem, 
power-series solution, Lie algebra, Lagrange-mesh method, asymptotic iteration method, generalized pseudo-spectral (GPS) 
method were invoked for their proper treatment. Many interesting aspects such as rearrangement and redistribution of ground 
and excited energy states, \emph{simultaneous} and \emph{incidental} degeneracy, change in hyperfine splitting constant as 
well as dipole shielding factor, nuclear magnetic screening constant, pressure, variation of static and dynamic polarizability, 
hyperpolarizability, information entropy, etc., were probed with varying confining radius $(r_c)$. A vast 
literature exists on the subject; here we refer to a selective set \cite{goldman92,aquino95,garza98,laughlin02,burrows06,
aquino07cha,baye08,ciftci09,montgomery12,cabrera13,roy15,solorzano16}. Eigenvalues and eigenfunctions of CHA can be solved 
\emph{exactly} in terms of Kummer M-function (confluent hypergeometric) \cite{burrows06}. 

In past twenty years, information measures were explored extensively for various quantum systems in both free and confinement
situations. Some such potentials are: P\"oschl-Teller \cite{sun2013quantum}, Rosen-Morse \cite{sun2013quantum1}, pseudo-harmonic 
\cite{yahya2015}, squared tangent well \cite{dong2014quantum}, hyperbolic \cite{valencia2015quantum}, position-dependent mass 
Schr\"odinger equation \cite{chinphysb,yanez2014quantum}, infinite circular well \cite{song2015shannon}, hyperbolic double-well 
(DW) potential \cite{sun2015shannon}, etc. Recently, entropic measures were successfully engaged to understand trapping and 
oscillation of a particle within symmetric, asymmetric DW potential \cite{neetik15,neetik16}, confined quantum harmonic 
oscillator \cite{ghosal16}, CHA \cite{aquino13,jiao17}, etc.
 
Information-theoretic measures like R\'{e}nyi entropy ($R$), Tsallis entropy ($T$), Shannon entropy ($S$) and Onicescu energy 
($E$), in atomic systems may provide detailed knowledge about diffusion of atomic orbitals, spread of electron density, 
periodic properties, correlation energy and so forth \cite{chatzisavvas05,romera08,grassi08,gallegos16a}. $R, T$, called 
information generating functionals, are directly connected to \emph{entropic moments} and completely quantify 
 density. Former 
has been effectively employed to illustrate quantum entanglement, chemical reactivity, de-coherence and localization properties 
of Rydberg states of atoms \cite{varga03,renner05,levay05,verstraete06,bialas06,salcedo09,liu15}. Similarly, $T$ has been 
implicated specially for non-extensive thermo-statistics \cite{tsallis04,naudts11} and gravitation \cite{plastino99,chen14}, 
etc. It is noteworthy that, $S, E$ are two special cases of $R, T$. Former measures extent of concentration of the system wave 
function in respective space, whereas latter symbolizes expectation values of density. $S$ has its application in 
illuminating colin conjecture, atomic avoided crossing, orbital free density functional theory \cite{nagy15,he15,site15,
alcoba16,gallegos16} in many-electron systems, etc. Likewise, $E$ has been widely used to estimate correlation energy and first 
ionisation potential \cite{gallegos16a}.    

In past few years appreciable attention has been paid to explore $S$ in both $r$ and $p$ space for CHA under soft and hard 
confinement \cite{aquino13}. Very recently, $S$ in conjugate space has been examined (for low-lying $s, p, d$ orbitals) with 
the help of variation principle employing Slater type orbitals \cite{jiao17}. However, rest of the information 
measures like $R, T, E$ have been attempted very rarely, with the exception of some first few $s$-states of CHA 
\cite{mukherjee17}. Hence, our primary motivation is to undertake a detailed analysis of $ R, T, S, E$ in a CHA-like system
in a systematic fashion for an arbitrary state characterized by principal and azimuthal quantum numbers $n,l$, in both spaces, 
with special emphasis on $l \neq 0$. Illustrative calculations are performed with \emph{exact} analytical wave functions in 
$r$-space; whereas in $p$ space, numerical wave functions are generated by executing Fourier transform on the eigenfunction 
of respective $r$-space orbitals. To put things in proper perspective, in this communication, $2p, 3d, 4f, 5g$ and $10s$-$10m$ 
states have been chosen as representatives. By considering all the acceptable $l$'s corresponding to a given $n$, one can follow 
the changes in behavior of $l$ states as the environment switches from \emph{free} to \emph{confinement}. Such a comparative study 
of these information measures are done with respect to their free Hydrogen atom (FHA) counterpart. We also inspect the nature of 
$R, T, S, E$ for hydrogenic isoelectronic series (by varying atomic number $Z$) inside the spherical impenetrable 
cavity, using the scaling properties \cite{patil07} satisfied by such a system. This time we restrict ourselves to ground 
state only; for extension to other states is straightforward. To this end, all measures in a CHA-like atoms are obtained 
in both $r$, $p$ and composite spaces. Note that such studies in a CHA are very rare and as already implied, most of the 
present results are offered here for the first time. Throughout the article, comparison with existing literature results are 
made wherever possible. Organization of this article is as follows. Section~II gives a brief account of the theoretical method 
used; Sec.~III presents a detailed discussion on the results of $R, T, S, E$ of CHA and H-isoelectronic series, while we 
conclude with a few remarks in Sec.~IV.

\section{Theoretical Method}
The time-independent, non-relativistic radial Schr\"odinger equation under the influence of confinement, without loss of 
generality, for a central potential, may be written as, 
\begin{equation}
\left[-\frac{1}{2} \ \frac{d^2}{dr^2} + \frac{\ell (\ell+1)} {2r^2} + v(r) +v_c (r) \right] \psi_{n,\ell}(r)=
\mathcal{E}_{n,\ell}\ \psi_{n,\ell}(r),
\end{equation}
where $v(r)=-\frac{Z}{r} \ (Z$ implies atomic number). Our desired effect of radial confinement inside an impenetrable hard 
cavity can be modeled by invoking the following form of potential: $v_c(r) = +\infty$ for $r > r_c$ and $0$ for $r \leq r_c$, 
where $r_c$ corresponds to radius of the cage. It is worthwhile mentioning that, atomic units are employed through out the 
calculations and $\rvec, \pvec$ subscripts denote quantities in full $r$ and $p$ spaces (including the angular part) 
respectively. 

On solving Eq.~(1) one can obtain following \emph{exact} form for eigenfunctions in CHA \cite{burrows06},
\begin{equation}
\psi_{n, l}(Zr)= N_{n, l}\left(2Zr\sqrt{-2\mathcal{E}_{n,l}}\right)^{l} \ _{1}F_{1}
\left[\left(l+1-\frac{1}{\sqrt{-2\mathcal{E}_{n,l}}}\right),(2l+2),2Zr\sqrt{-2\mathcal{E}_{n,l}}\right] 
e^{-Zr\sqrt{-2\mathcal{E}_{n,l}}}.
\end{equation}
Here, $N_{n, l}$ represents the normalization constant, $\mathcal{E}_{n,l}$ prevails to energy of a given $n,l$ state while 
$_1F_1\left[a,b,r\right]$ refers to confluent hypergeometric function. Allowed energies at a given $r_c$ can be retrieved by 
applying the Dirichlet boundary condition that $\psi_{n,\ell} (0)= \psi_{n,\ell} \ (r_c)=0$ and finding the zeros of 
$_{1}F_{1}$, such that, 
\begin{equation}
_{1}F_{1}\left[\left(l+1-\frac{1}{\sqrt{-2\mathcal{E}_{n,l}}}\right),(2l+2),2Zr_{c}\sqrt{-2\mathcal{E}_{n,l}}\right]=0. 
\end{equation}

For a particular $l$, first root signifies energy of the lowest state having $(n_{lowest}=l+1)$, and consecutive roots 
imply excited states. Note that, for construction of exact wave function of CHA for a specific state, one needs 
to provide the energy eigenvalue of that state. In our present calculation, $\mathcal{E}_{n,l}$ of CHA are computed by invoking 
the GPS \cite{roy15} method. This is applied, because in this pursuit, we are interested in the information measures in CHA, 
for which GPS energies have been found to be sufficiently accurate to provide correct eigenvalues and eigenfunctions. Over the 
years, this has been tested in a varied case of important model and realistic potentials, including both \emph{free and 
confinement} situations \cite{roy04,sen06,roy13,roy15}. Also, it is obvious from Eq.~(2) that, $\mathcal{E}_{n,l}$ depends on 
the product $Zr_c$. Hence in spite of changes in $Z$ and $r_c$ separately, if their product remains constant, then 
$\mathcal{E}_{n,l}$ will not be affected.  
 
The angular part has following common form in both $r$ and $p$ spaces ($P_{l}^{m} (\cos \theta)$ signifies usual associated 
Legendre polynomial), 
\begin{equation}
Y_{l,m} ({\Omega}) =\Theta_{l,m}(\theta) \ \Phi_m (\phi) = (-1)^{m} \sqrt{\frac{2l+1}{4\pi}\frac{(l-m)!}{(l+m)!}} 
\ P_{l}^{m}(\cos \theta)\ e^{-im \phi}.  
\end{equation}
The $p$-space wave function ($\pvec = \{ p, \Omega \}$) for a particle in a central potential is obtained from respective 
Fourier transform of its $r$-space counterpart, and as such, is given below,
\begin{equation}
\begin{aligned}
\psi_{n,l}(p) & = & \frac{1}{(2\pi)^{\frac{3}{2}}} \  \int_0^{r_c} \int_0^\pi \int_0^{2\pi} \psi_{n,l}(r) \ \Theta(\theta) 
 \Phi(\phi) \ e^{ipr \cos \theta}  r^2 \sin \theta \ \mathrm{d}r \mathrm{d} \theta \mathrm{d} \phi,  \\
      & = & \frac{1}{2\pi} \sqrt{\frac{2l+1}{2}} \int_0^{r_c} \int_0^\pi \psi_{n,l} (r) \  P_{l}^{0}(\cos \theta) \ 
e^{ipr \cos \theta} \ r^2 \sin \theta  \ \mathrm{d}r \mathrm{d} \theta.  
\end{aligned}
\end{equation}
Note that, here $\psi(p)_{n,l}$ needs to be normalized. Integrating over $\theta$ and $\phi$ variables, Eq.~(5) may be 
further rewritten as given below,  
\begin{equation}
\psi_{n,l}(p)=(-i)^{l} \int_0^{r_c} \  \frac{\psi_{n,l}(r)}{p} \ f(r,p)\mathrm{d}r.    
\end{equation}
Depending on $l$, $f(r,p)$ can be expressed in following simplified form ($m'$ starts with 0),  
\begin{equation}
\begin{aligned}
f(r,p) & = & \sum_{k=2m^{\prime}+1}^{m^{\prime}<\frac{l}{2}} a_{k} \ \frac{\cos pr}{p^{k}r^{k-1}} +  
            \sum_{j=2m^{\prime}}^{m^{\prime}=\frac{l}{2}} b_{j} \ \frac{\sin pr}{p^{j}r^{j-1}}, \ \ \ \ \mathrm{for} \ 
            \mathrm{even} \ l,   \\
f(r,p) & = & \sum_{k=2m^{\prime}}^{m^{\prime}=\frac{l-1}{2}} a_{k} \ \frac{\cos pr}{p^{k}r^{k-1}} +  
\sum_{j=2m^{\prime}+1}^{m^{\prime}=\frac{l-1}{2}} b_{j} \ \frac{\sin pr}{p^{j}rerst_cha_6.tex^{j-1}}, \ \ \ \ \mathrm{for} \ \mathrm{odd} \ l.
\end{aligned} 
\end{equation}
The coefficients $a_{k}$, $b_{j}$ of even-$l$ and odd-$l$ states are calculated using Eq.~(5).

Let $\rho(\rvec)$ and $\Pi(\pvec)$ denote normalized position and momentum electron densities for CHA. Then, position, momentum shannon entropies ($S_{\rvec}, 
S_{\pvec}$) and their sum ($S_{t}$) for H-like atoms are defined in terms of expectation values of logarithmic probability density functions 
and $Z$, which for a central potential further simplify to, 
\begin{equation}
\begin{aligned}
S_{\rvec}(Z) & = -3 \ln Z + S_{\rvec}(Z=1), \ \ \ S_{\pvec}(Z) = 3 \ln Z + S_{\pvec}(Z=1) \\ 
S_{\rvec}(Z=1) & =  -\int_{{\mathcal{R}}^3} \rho(\rvec) \ \ln [\rho(\rvec)] \ \mathrm{d} \rvec ,  \ \ \
S_{\pvec}(Z=1) =  -\int_{{\mathcal{R}}^3} \Pi(\pvec) \ \ln [\Pi(\pvec)] \ \mathrm{d} \pvec     \\
S_{t} & = S_{\rvec}(Z)+S_{\pvec}(Z) = S_{\rvec}(Z=1)+S_{\pvec}(Z=1)
\end{aligned} 
\end{equation}
The above equation clearly suggests that, at a fixed $r_c$, both $S_{\rvec}(Z)$ and $S_{\pvec}(Z)$ are linear functions of 
logarithm of $Z$ with slope $-3$ and $3$ respectively. Moreover, $S_{\rvec}(Z=1)$, $S_{\pvec}(Z=1)$ act as intercepts in $r$-
and $p$-space equations. The last equation implies that addition of $S_{\rvec}(Z)$ and $S_{\pvec}(Z)$ produces same result
as one obtains from the corresponding sum for $Z=1$. Hence $S_{t}$ will remain unaltered with change in $Z$. Actually 
it has been established \cite{patil07} that it solely depends on $Zr_c$ product as evident from Eq.~(3), rather than 
individual $Z$ and $r_c$. Here and in the discussion throughout the article, $Z$ dependence is identified in the parentheses 
of all respective measures, for $Z > 1$. For CHA ($Z=1$), there are no parentheses in the expressions of $R,T,S,E$.

Similarly, R{\'e}nyi entropies of order $\lambda (\neq 1)$ are obtained by taking logarithm of $\lambda$-order entropic moment. 
In spherical polar coordinate, they are expressed as, 
\begin{equation}
\begin{aligned} 
R_{\rvec}^{\lambda}(Z=1)  =  \frac{1}{1-\lambda} \ln \left(\int_{{\mathcal{R}}^3} \rho^{\lambda}(\rvec)\mathrm{d} \rvec \right)  = &
\frac{1}{(1-\lambda)} \ln \left(2\pi\int_0^\infty [\rho(r)]^{\lambda} r^2 \mathrm{d}r \int_0^\pi [\chi(\theta)]^{\lambda} \sin \theta \mathrm{d}\theta \right) \\ 
 = & \frac{1}{(1-\lambda)}\left( \ln 2\pi + \ln [\omega^{\lambda}_r] + \ln [\omega^{\lambda}_{(\theta, \phi)}] \right),  \\
R_{\pvec}^{\lambda}(Z=1)  =  \frac{1}{1-\lambda} \ln \left[\int_{{\mathcal{R}}^3} \Pi^{\lambda}(\pvec)\mathrm{d} \pvec \right]  = &
\frac{1}{(1-\lambda)} \ln \left(2\pi \int_{0}^\infty [\Pi(p)]^{\lambda} p^2 \mathrm{d}p \int_0^\pi [\chi(\theta)]^{\lambda} \sin \theta \mathrm{d}\theta \right) \\
 = & \frac{1}{(1-\lambda)}\left( \ln 2\pi + \ln [\omega^{\lambda}_p] + \ln [\omega^{\lambda}_{(\theta, \phi)}] \right).
\end{aligned} 
\end{equation}          
where $\omega^{\lambda}_{\tau}$'s are entropic moments in $\tau$ ($r$ or $p$ or $\theta, \phi$) space with order 
$\lambda$, having forms,
\begin{equation}                                                                                                 
\omega^{\lambda}_r= \int_0^\infty [\rho(r)]^{\lambda} r^2 \mathrm{d}r, \ \ \   erst_cha_6.tex                                  
\omega^{\lambda}_p= \int_{0}^\infty [\Pi(p)]^{\lambda} p^2 \mathrm{d}p, \ \ \                                    
\omega^{\lambda}_{(\theta, \phi)}= \int_0^\pi [\chi(\theta)]^{\lambda} \sin \theta \mathrm{d}\theta.             
\end{equation}                                                                                                   
Here if $\lambda$ corresponds to $\alpha$, $\beta$ in $r$, $p$ spaces respectively, then they are related as     
$\frac{1}{\alpha}+\frac{1}{\beta}=2.$ In that case, one can define R{\'e}nyi entropy sum as,                             
\begin{equation}                                                                                                 
%\begin{split}                                                                                                    
R_{t}^{(\alpha,\beta)}  =  \frac{2-\alpha-\beta}{(1-\alpha)(1-\beta)} \ \ln 2\pi+ \frac{1}{(1-\alpha)} 
\left( \ln [\omega^{\alpha}_r]+ \ln [\omega^{\alpha}_{(\theta, \phi)}]\right)  
 + \frac{1}{(1-\beta)}\left( \ln [\omega^{\beta}_p]+ \ln [\omega^{\beta}_{(\theta, \phi)}]\right). 
%\end{split} 
\end{equation}
Equation~(9) suggests that, at a particular $r_c$, like $S_{\rvec}(Z)$, $S_{\pvec}(Z)$, both $R_{\rvec}^{\lambda}(Z)$ and 
$R_{\pvec}^{\lambda}(Z)$ also linearly depend on $\ln Z$ (slope $-3, 3$ respectively) with intercepts $R_{\rvec}^{\lambda}(Z=1)$ 
and $R_{\pvec}^{\lambda}(Z=1)$ respectively. Further, as with $S_{t}$, $R^{(\alpha, \beta)}$ also remains 
unimpacted with change of $Z$.

In a similar fashion, Tsallis entropies \cite{tsallis88} in $r, p$ space and their product can be written down as below, 
\begin{equation}
\begin{aligned}
T_{\rvec}^{\alpha} (Z)= \left(\frac{1}{\alpha-1}\right)\left[1-\frac{1}{Z^3}\int_{{\mathcal{R}}^3} \rho^{\alpha}(\rvec)\mathrm{d} 
\rvec \right] 
    & = & \left(\frac{1}{\alpha-1}\right)\left[1-\frac{1}{Z^3}\omega^{\alpha}_r \omega^{\alpha}_{(\theta, \phi)}\right] \\
T_{\pvec}^{\beta} (Z)= \left(\frac{1}{\beta-1}\right)\left[1-Z^{3}\int_{{\mathcal{R}}^3} \Pi^{\beta}(\pvec)\mathrm{d} \pvec 
\right] 
    & = & \left(\frac{1}{\beta-1}\right)\left[1-Z^{3}\omega^{\beta}_r \omega^{\beta}_{(\theta, \phi)}\right] \\
T_{t}^{(\alpha,\beta)}(Z)= T_{\rvec}^{\alpha} (Z) T_{\pvec}^{\beta} (Z)
\end{aligned}
\end{equation}
One sees that, $T_{\rvec}^{\alpha} (Z)$ reduces and $T_{\pvec}^{\beta} (Z)$ enhances with rise of $Z$. Note that 
$\alpha=\beta= 2$ leads to Onicescu energy $E$, which in $r$ and $p$ spaces are given as, 
\begin{equation}
\begin{aligned}
E_{\rvec} (Z)= Z^3\omega^{2}_r \omega^{2}_{(\theta, \phi)}= Z^3 E_{\rvec} (Z=1) \\
E_{\pvec} (Z)= \frac{1}{Z^{3}}\omega^{2}_r \omega^{2}_{(\theta, \phi)} = \frac{1}{Z^3} E_{\pvec} (Z=1) \\
E_{t} (Z)= E_{\rvec} (Z) E_{\pvec} (Z) = E_{\rvec} (Z=1) E_{\pvec} (Z=1).
\end{aligned}
\end{equation}
Here, $E_{t}$ represents onicescu energy product.
Dependence of $E_{\rvec} (Z)$ and $E_{\pvec} (Z)$ on $Z$ is seen to be opposite to that of the previous three measures 
discussed above. Thus $E_{\rvec} (Z)$ grows up and $E_{\pvec} (Z)$ falls off with rise of $Z$. But, as usual, $E_{t}(Z)$ 
remains unchanged as $Z$ modifies.

\begingroup            %%% table I
\squeezetable
\begin{table}
\caption{R\'enyi entropies, $R_{\rvec}^{\alpha},~R_{\pvec}^{\beta}$ and $R_{t}^{(\alpha,\beta)}$ for $2p,~3d$ states of 
CHA at selected $r_c$ values, with $\alpha= \frac{3}{5}$ and $\beta= 3$. See text for details.}
\centering
\begin{ruledtabular}
\begin{tabular}{llll|llll}
$r_c$  &  $R_{\rvec}^{\alpha}$ & $R_{\pvec}^{\beta}$ & $R{t}^{(\alpha,\beta)}$ &   $r_c$ &   $R_{\rvec}^{\alpha} $  &    
$R_{\pvec}^{\beta} $    & $R_{t}^{(\alpha,\beta)}$   \\
\hline
\multicolumn{4}{c}{$2p^{!}$}    \vline  &      \multicolumn{4}{c}{$3d^{\P}$}    \\
\hline
0.1  &  $-$6.16888358521  & 12.8086549   & 6.6397714    & 0.1   &  $-$6.11864461237   &     13.2299266     & 7.111282       \\
0.2  &  $-$4.09121892459  & 10.7314035   & 6.6401846    & 0.2   &  $-$4.03973495813   &     11.1514674     & 7.1117325      \\
0.3  &  $-$2.87663035551  &  9.5172358   & 6.6406055    & 0.3   &  $-$2.82387786690   &      9.9360644     & 7.1121866      \\
0.5  &  $-$1.34785892643  &  7.9893299   & 6.641471     & 0.5   &  $-$1.29249680384   &      8.4056025     & 7.1131057     \\
0.8  &   0.05635500373    &  6.5864754   & 6.6428304    & 0.8   &   0.11582086058   &      6.9986916     & 7.1145125       \\
1    &   0.72175456102    &  5.9220245   & 6.643779     & 1     &   0.78408877031   &      6.3313805     & 7.1154693       \\
5    &   5.43150144784    &  1.2403914   & 6.6718928    & 7.5   &   6.77107758150   &      0.3857834     & 7.1568610       \\
15   &   7.78024380659    & $-$0.924233  & 6.8560108    & 15    &   8.69500260562   &    $-$1.45238070   & 7.2426219       \\
25   &   7.92394945000    & $-$0.9742317 & 6.9497178    & 50    &   9.92580433711   &    $-$2.311277606  & 7.614526731    \\
45   &   7.92577664624    & $-$0.9736503 & 6.9521263    & 100   &   9.92600859049   &    $-$2.311283603  & 7.614724988    \\  
\end{tabular}
\end{ruledtabular}
\begin{tabbing}
$^!$$R^{\alpha}_{\rvec},~R^{\beta}_{\pvec},~R_{t}^{(\alpha, \beta)}$ in FHA for $2p$ states ($|m|=0$) are: 7.925776675482,~$-$0.9736503771629,~6.952126298319. \\ 
$^\P$$R^{\alpha}_{\rvec},~R^{\beta}_{\pvec},~R_{t}^{(\alpha, \beta)}$ in FHA for $3d$ states ($|m|=0$) are: 9.926008594642,~$-$2.311283609195,~7.614724985446. 
\end{tabbing}
\end{table}
\endgroup

\section{Result and Discussion}
At the outset, it is appropriate to mention a few things regarding the presentation. Here, our focus is to uncover the impact 
of an impenetrable spherical cage on \emph{non-zero} $l$ states of CHA by using information-theoretic measures. The \emph{net} 
information measures in conjugate $r$ and $p$ space of CHA may be segmented into two separate contributions, \emph{viz.}, (i) 
a radial and (ii) an angular part. In CHA, the radial barrier changes from infinity to a finite region without affecting 
angular boundary conditions. So angular portion remains invariant in $r$ and $p$ spaces; moreover they will also 
not change with respect to boundary condition in $r_c$ in a CHA. However, they will be affected by $l$, $m$ quantum numbers. 
In current calculation, we have kept magnetic quantum number $m$ fixed at 0, unless stated otherwise. It is clear from Eq.~(2) 
that, in $r$ space, radial wave functions are available in closed analytical form. In $p$ space, numerical wave functions are 
achieved by employing Fourier transform on respective $r$-space eigenfunctions. All our results provided in tables and figures 
are computed numerically. It is expected that, a gradual increase in $r_{c}$ should lead to a delocalization in the system in 
such a fashion that, when $r_{c} \rightarrow \infty$, it should unfold to FHA. On the contrary, when $r_{c} \rightarrow 0$, 
impression of confinement is maximum. Thus, it is convenient to explore our analysis by choosing some specific $r_{c}$ 
values in the range of $0.1$ to $100$. This parametric rise in $r_{c}$ reveals evolution of system from maximum 
confinement to a free system. It may be remarked that, a detailed systematic analysis of these measures in low-lying $s$ 
states, along the lines of current work, has been initiated by present authors and will be published elsewhere 
\cite{mukherjee17}. 

\begingroup            %%% table II
\squeezetable
\begin{table}
\caption{$R_{\rvec}^{\alpha}$ and $R_{\pvec}^{\beta}$ for $10l$ states of CHA at seven different $r_c$ values, namely 
$0.1, 0.5, 1, 10, 40, 80, 100$. Here, $\alpha, \beta$ are chosen to be $\frac{3}{5}$ and 3 respectively. See text for details.}
\centering
\begin{ruledtabular}
\begin{tabular}{l|lllllll}
 \multicolumn{8}{c}{$R_{\rvec}^{\alpha}$}    \\
\hline
$l$  &  $r_c=0.1$ & $r_c=0.5$  & $r_c=1$  &   $r_c=10$  &   $r_c=40$   &   $r_c=80$  & $r_c=100$   \\
\hline
0    &    $-$6.0792476535 & $-$1.2500151671  & 0.8305932083  &  7.7616144459   & 11.9866555713  &  14.1264898722  & 14.8227228363   \\
1    &    $-$6.3891801965 & $-$1.5604409325  & 0.5195411162  &  7.4386341260   & 11.6485744296  &  13.7938994865  & 14.4924250904   \\
2    &    $-$6.3886059693 & $-$1.5600883216  & 0.5196127247  &  7.4330589592   & 11.6242072771  &  13.7633595572  & 14.4630426402   \\
3    &    $-$6.3665110491 & $-$1.5381245479  & 0.5414121420  &  7.4515237538   & 11.6283879357  &  13.7541478868  & 14.4519363360   \\
4    &    $-$6.3359608527 & $-$1.5076548776  & 0.5717793565  &  7.4797957142   & 11.6465100421  &  13.7573382983  & 14.4505385080  \\
5    &    $-$6.2994173707 & $-$1.4711571510  & 0.6082191748  &  7.5150095855   & 11.6749072268  &  13.7721369041  & 14.4592392391   \\
6    &    $-$6.2581496967 & $-$1.4299022978  & 0.6494567192  &  7.5558021952   & 11.7119190030  &  13.7982974268  & 14.4788958301   \\
7    &    $-$6.2143631650 & $-$1.3860901364  & 0.6933001405  &  7.6001063439   & 11.7558856869  &  13.8354493928  & 14.5099982856   \\
8    &    $-$6.1748354691 & $-$1.3464778435  & 0.7330180632  &  7.6416699182   & 11.8023199187  &  13.8818610231  & 14.5518827831   \\
9    &    $-$6.1702611858 & $-$1.3416902256  & 0.7380727967  &  7.6515945538   & 11.8288561033  &  13.9252706635  & 14.5938547362  \\
\hline
\multicolumn{8}{c}{$R_{\pvec}^{\beta}$}    \\
\hline
0    &    17.764042 &    12.938472  & 10.863484  &  4.225406     & $-$0.104732   & $-$2.194616   & $-$3.015142  \\
1    &    16.990848 &    12.164511  & 10.087976  &  3.319124     & $-$0.643120   & $-$2.666435   & $-$3.423219  \\
2    &    16.609713 &    11.782989  &  9.705776  &  2.886201     & $-$0.749884   & $-$2.917930   & $-$3.674227  \\
3    &    16.31713  &    11.490168  &  9.412558  &  2.566775     & $-$1.040808   & $-$3.221412   & $-$3.822730  \\
4    &    16.059368 &    11.2322271 &  9.154340  &  2.292588     & $-$1.439844   & $-$3.385452   & $-$4.054590  \\
5    &    15.8154003 &   10.9881122 &  8.910008  &  2.037463     & $-$1.820272   & $-$3.487284   & $-$4.250016  \\
6    &    15.573099 &    10.745675  &  8.667378  &  1.786891     & $-$2.166955   & $-$3.735834   & $-$4.383047  \\
7    &    15.322708 &    10.495141  &  8.416651  &  1.529727     & $-$2.494875   & $-$4.1772092  & $-$4.645285  \\
8    &    15.052818 &    10.225072  &  8.146348  &  1.253299     & $-$2.824948   & $-$4.672206   & $-$5.156688  \\
9    &    14.7302406 &    9.9021894 &  7.823079  &  0.9218974    & $-$3.204373   & $-$5.196718   & $-$5.787983  \\
\end{tabular}
\end{ruledtabular}
\end{table}
\endgroup

Our presentation strategy is as follows. Initially, $2p, 3d, 4f, 5g$ states are selected for analysis of various information 
measures; they all individually represent the lowest (nodeless) state of respective $l$. This will help us follow changes in 
$R,S,T,E$ with respect to alterations in $l$. Additionally we also explore all the $l$ states corresponding to a given $n$ 
(here chosen 10) to understand the outcome as the count of radial nodes vary. Actually, for a given $l$, an increase in $n$ 
enhances spreading as well as number of radial nodes, whereas an increment in $l$ within in a given $n$ reduces radial node.  

\begin{figure}                         %%%Fig. 1, FHA
\begin{minipage}[c]{0.4\textwidth}\centering
\includegraphics[scale=0.75]{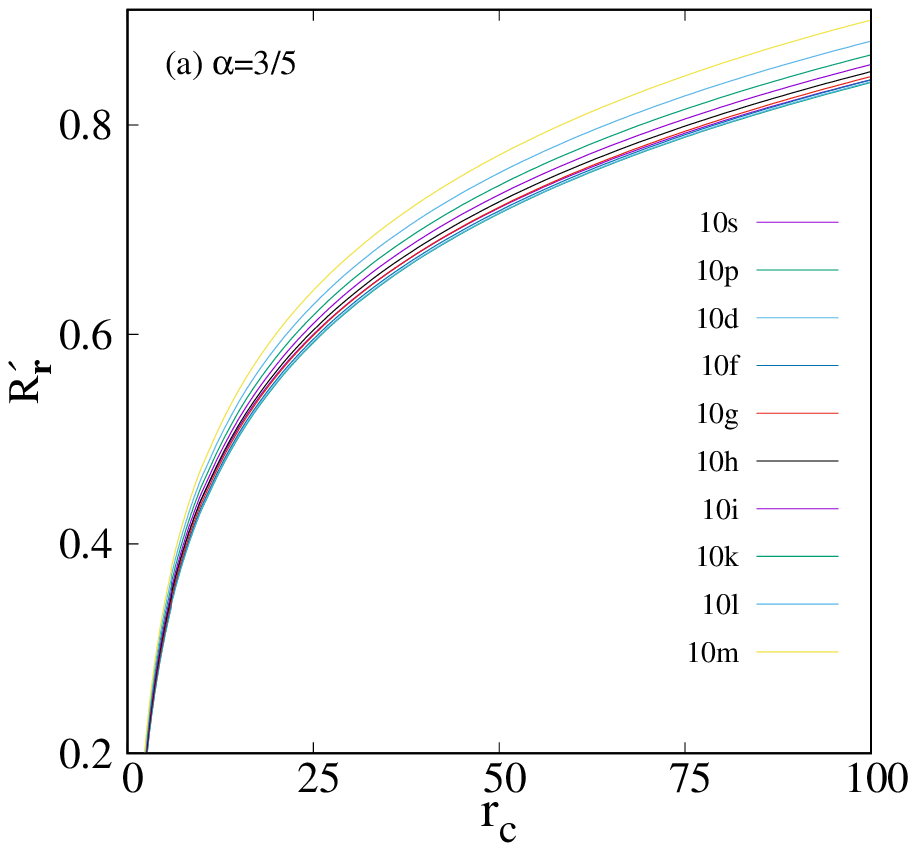}
\end{minipage}%
\hspace{0.1in}
\begin{minipage}[c]{0.5\textwidth}\centering
\includegraphics[scale=0.75]{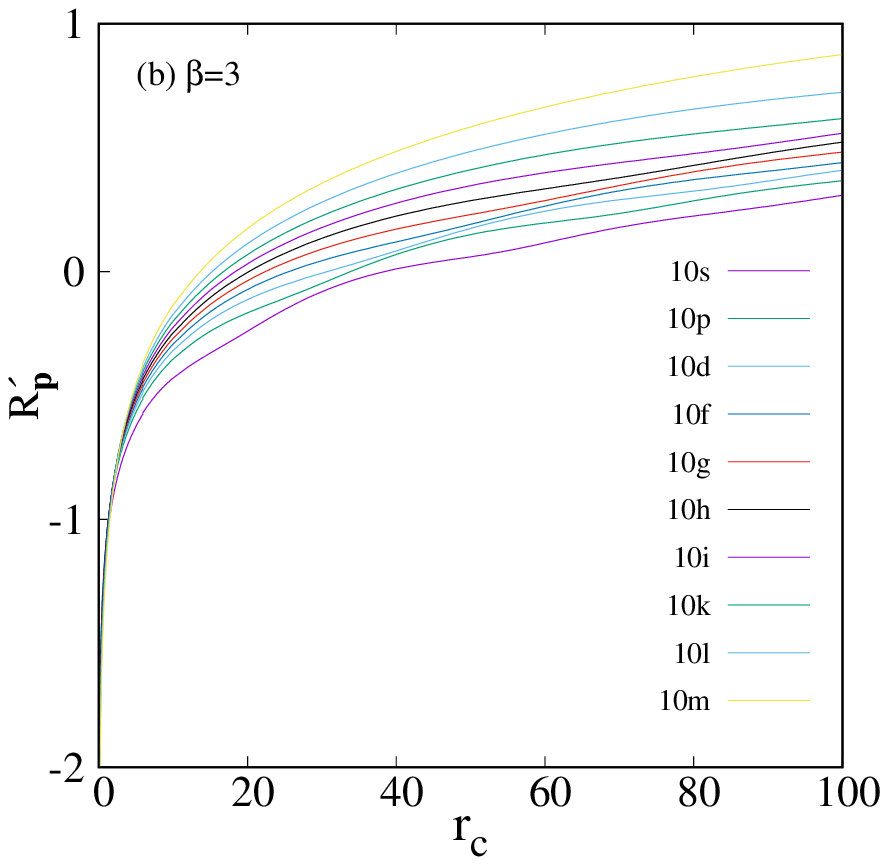}
\end{minipage}%
\caption{Variation of $R_{\rvec}^{\prime}$, $R_{\pvec}'$ with $r_c$ for $10s$--$10m$ states of CHA in panels (a), (b). 
$\alpha, \beta $ are chosen to be $\frac{3}{5}$ and 3 respectively. See text for detail.}
\end{figure}

To begin with, Table~I displays our calculated $R_{\rvec}^{\alpha}, R_{\rvec}^{\alpha}$, $R_{t}^{(\alpha, \beta)}$ for 
$2p, 3d$ states of CHA at a selected set of $r_c$; which differ from state to state. Similarly, Table S1 in supplementary material (SM) portrays all the above quantities
as a function of $r_c$ for $4f, 5g$ states. In this and all following tables related to these four states, information quantities are provided 
at \emph{same} set including ten $r_c$'s. In all occasions, $R_{\rvec}^{\alpha}$'s progress continuously with $r_c$ and finally converge to 
respective FHA behavior after some larger finite $r_c$. Interestingly, for all states, $R_{\rvec}^{\alpha}$ remain negative up to $r_c=0.5$, 
changing sign after that. In contrast, $R_{\pvec}^{\beta}$'s, tend to diminish with rise of $r_c$; eventually they also merge to FHA (negative for all) 
in the end. As a combination of these two effects, $R^{(\alpha, \beta)}$'s steadily grow with progress of $r_c$ and 
as usual they coalesce to respective borderline values finally. It is worth mentioning that, at \emph{small} $r_c$, R\'enyi
entropies in $r, p$ space obey same order: $R_{\rvec}^{\alpha}(5g)>R_{\rvec}^{\alpha}(4f)>R_{\rvec}^{\alpha}(3d) 
>R_{\rvec}^{\alpha}(2p)$ and $R_{\pvec}^{\beta}(5g)>R_{\pvec}^{\beta}(4f)>R_{\pvec}^{\beta}(3d)>R_{\pvec}^{\beta}(2p)$. 
But, at $r_c \rightarrow \infty$ limit, ordering in $p$ space completely reverses, while the $r$-space ordering is 
retained. The distribution pattern in $r$ space is observed presumably because, an increase in both $n, l$ leads to more 
delocalization of CHA. Next Table~II provides $R_{\rvec}^{\alpha}, R_{\pvec}^{\beta}$ for all states having $l=0-9$, within 
$n=10$ of CHA, at seven selected $r_c$, namely $0.1, 0.5, 1, 10, 40, 80, 100$, featuring strong to medium confinement. At smaller 
$r_c$ ($<1$) region, $R_{\rvec}^{\alpha}$'s remain ($-$)ve for all ten $l$; for rest five $r_c$ ($>1$), the sign reverses. 
Numerical values surge with $r_c$ for all $l$. In all seven instances of $r_c$, $R_{\rvec}^{\alpha}$'s initially fall as 
$l$ increments, then attain a minimum for some $l$ and again go up. {\color{red} In first three $r_c$ ($0.1, 0.5, 1$) these minima 
occur at $l=1$. At $r_c= 10,~40$ these happen at $l=2$. Further, These minima shift to $l=3$ and $l=4$ for $r_c=80$ and $100$ respectively.}  
In contrast, numerical values of $R_{\pvec}^{\beta}$ gradually lower with growth of $r_c$ for each $l$. Furthermore, in all seven $r_c$, it falls down with 
$l$ without going through any inflection points unlike its corresponding $r$-space counterpart. To the best of our knowledge, 
no such results have been reported for R\'enyi entropy in CHA. Hence they could not be directly compared, and hopefully 
would provide useful guidelines for future work. 

\begingroup        %%% table III    
\squeezetable
\begin{table}
\caption{Tsallis entropies $T_{\rvec}^{\alpha},~T_{\pvec}^{\beta}$ and $T^{(\alpha,\beta)}$ for $2p,~3d$ states 
of CHA at various $r_c$, for $\alpha$, $\beta$ as $\frac{3}{5}$ and 3 respectively. For more details, see text.}
\centering
\begin{ruledtabular}
\begin{tabular}{llll|llll}
$r_c$  &  $T_{\rvec}^{\alpha}$ & $T_{\pvec}^{\beta}$  & $T_{t}^{(\alpha,\beta)} $ &   $r_c$ &   $T_{\rvec}^{\alpha} $  
       &    $T_{\pvec}^{\beta} $    & $T_{t}^{(\alpha,\beta)} $  \\
\hline
\multicolumn{4}{c}{$2p^!$}    \vline  &      \multicolumn{4}{c}{$3d^\P$}    \\
\hline
0.1 & $-$2.2880198630 & 0.4999999999 & $-$1.1440099313 & 0.1 & $-$2.28371690687 &  0.499999999998  & $-$1.1418584534    \\
0.2 & $-$2.0133435459 & 0.4999999997 & $-$1.0066717723 & 0.2 &  $-$2.00321763758 &  0.499999999896  & $-$1.0016088186   \\
0.3 & $-$1.7089241387 & 0.4999999972 & $-$0.8544620646 & 0.3 &  $-$1.69205429127 &  0.499999998828  & $-$0.8460271437   \\
0.5 & $-$1.0418811277 & 0.4999999425 & $-$0.5209405039 & 0.5 &  $-$1.00923112502 &  0.499999974999  & $-$0.5046155373   \\
0.8 & 0.0569949807  & 0.4999990493 & 0.0284974362     & 0.8   &   0.11854567062  &  0.499999583146  & 0.0592727859     \\
1   & 0.8367342402  & 0.4999964094 & 0.4183641157     & 1     &   0.92097719569  &  0.499998416555  & 0.4604871395      \\
5   & 19.452725733  & 0.4581611514 & 8.9124832214     & 7.5   &   35.0142839008  &  0.268855906272  & 9.4137970306     \\
15  & 53.670306013  & $-$2.6750355 & $-$143.5699738   & 15    &   78.4872531785  &  $-$8.630443042  & $-$677.379768    \\
25  & 56.993705192  & $-$3.0089477 & $-$171.4910836   & 50    &  130.0039513800  &  $-$50.37685125  & $-$6549.18972146  \\
45  & 57.037203756  & $-$3.0048699 & $-$171.38938035    & 100   &  130.0147775743   &  $-$50.37746147  & $-$6549.81444873         \\
\end{tabular}
\end{ruledtabular}
\begin{tabbing}
$^!$$T^{\alpha}_{\rvec},~T^{\beta}_{\pvec},~T^{(\alpha, \beta)}$ in FHA for $2p$ states ($|m|=0$) are: 57.037204453034,$-$3.0048705041661,$-$171.38941330101483. \\ 
$^\P$$T^{\alpha}_{\rvec},~T^{\beta}_{\pvec},~T^{(\alpha, \beta)}$ in FHA for $3d$ states ($|m|=0$) are: 130.014775913339,$-$50.377462107980,$-$6549.814447051782. 
\end{tabbing}
\end{table}
\endgroup

\begingroup           %%% table IV 
\squeezetable
\begin{table}
\caption{Tsallis entropies $T_{\rvec}^{\alpha}$ and $T_{\pvec}^{\beta}$ for all $l$ states corresponding to $n=10$, of CHA at 
various $r_c$, for $\alpha$ and $\beta$ as $\frac{3}{5}$ and 3 respectively. More details can be found in the text.}
\centering
\begin{ruledtabular}
\begin{tabular}{l|lllllll}
$l$  &  $r_c=0.1$ & $r_c=0.5$  & $r_c=1$  &   $r_c=10$  &   $r_c=40$   &   $r_c=80$   &  $r_c=100$     \\
\hline
\multicolumn{8}{c}{$T_{\rvec}^{\alpha}$}    \\
\hline
0    &    $-$2.28028155100 & $-$0.9836825500  & 0.9852090051 &  53.2532949166 & 299.6588765682  & 708.652348746    & 937.0301091469   \\
1    &    $-$2.30589991419 & $-$1.1607438353  & 0.5774679920 &  46.4962802663 & 261.4396566474  & 620.0665476461   & 820.75069711    \\
2    &    $-$2.30585532605 & $-$1.1605549274  & 0.5775561424 &  46.3871370347 & 258.8795701728  & 612.5075565192   & 811.1316848206  \\
3    &    $-$2.30413187694 & $-$1.1487353757  & 0.6045090554 &  46.7495501415 & 259.3170313452  & 610.2456275455   & 807.5251254619  \\
4    &    $-$2.30172366881 & $-$1.1321655705  & 0.6424491328 &  47.3096638083 & 261.2217971179  & 611.0280909941   & 807.072341736  \\
5    &    $-$2.29880409954 & $-$1.1120499540  & 0.6885886895 &  48.0162247111 & 264.2344575738  & 614.6706053311   & 809.8947988758  \\
6    &    $-$2.29545538209 & $-$1.0889560609  & 0.7416206960 &  48.8472617617 & 268.2127601388  & 621.1627157435   & 816.3075412867 \\
7    &    $-$2.29184130513 & $-$1.0640097551  & 0.7989716149 &  49.7653312815 & 273.0158081039  & 630.500042316    & 826.5579416463 \\
8    &    $-$2.28852393521 & $-$1.0410753919  & 0.8518014821 &  50.6415282666 & 278.1809736722  & 642.3612266162   & 840.5648183275 \\
9    &    $-$2.28813664043 & $-$1.0382788056  & 0.8585853232 &  50.8529116901 & 281.1761223672  & 653.6562832778   & 854.8383313138 \\
\hline
\multicolumn{8}{c}{$T_{\pvec}^{\beta}$}    \\
\hline
0    &    0.4999999999999998 & 0.49999999999 & 0.4999999998  & 0.49989313660   & $-$0.116508     & $-$39.789255    &  $-$207.416554    \\
1    &    0.4999999999999991 & 0.49999999998 & 0.4999999991  & 0.49934534041   & $-$1.309576     & $-$103.015651   &  $-$469.762391    \\
2    &    0.4999999999999981 & 0.49999999997 & 0.4999999981  & 0.49844386357   & $-$1.740324     & $-$170.679518   &  $-$776.396253    \\
3    &    0.4999999999999966 & 0.49999999994 & 0.4999999966  & 0.49705220303   & $-$3.508707     & $-$313.589136   &  $-$1045.065137   \\
4    &    0.4999999999999944 & 0.49999999991 & 0.4999999944  & 0.49489902295   & $-$8.404357     & $-$435.549958   &  $-$1661.925267   \\
5    &    0.4999999999999908 & 0.49999999985 & 0.4999999908  & 0.49150326392   & $-$18.556282    & $-$534.047620   &  $-$2456.963057   \\
6    &    0.4999999999999851 & 0.49999999976 & 0.4999999851  & 0.48597521532   & $-$37.620905    & $-$878.267905   &  $-$3206.037005   \\
7    &    0.4999999999999754 & 0.49999999961 & 0.4999999755  & 0.47654334855   & $-$72.949846    & $-$2123.956362  &  $-$5417.179246   \\
8    &    0.4999999999999579 & 0.49999999934 & 0.4999999580  & 0.45922740769   & $-$141.630950   & $-$5716.873607  &  $-$15065.997051  \\
9    &    0.4999999999999197 & 0.49999999874 & 0.4999999198  & 0.42089205609   & $-$303.065930   & $-$16321.820720 &  $-$53253.000550  \\
\end{tabular}
\end{ruledtabular}
\end{table}
\endgroup

Above changes of Table~II are depicted in Fig.~1 in the form of ratios $R_{\rvec}'=\left(\frac{R_{\rvec}^{\alpha} (CHA)}
{R_{\rvec}^{\alpha}(FHA)}\right)$ and $R_{\pvec}'=\left(\frac{R_{\pvec}^{\beta} (CHA)}{R_{\pvec}^{\beta}(FHA)}\right)$ for 
all $l$ states of $n=10$. For convenience, R\'enyi entropies of CHA, in both spaces, in this occasion, are divided by their 
respective FHA values. First it is relevant to recall the following facts about these two relative quantities. For any given 
$n, l$ $R_{\rvec}^{\alpha}$ enhances and $R_{\pvec}^{\beta}$ declines with rise of $r_c$, whereas for a FHA, they assume 
maximum and minimum values. Further, this minimum in the latter has negative sign. Such a division by their respective 
free counterpart makes these quantities unit-less and keeps upper bound to unity. This facilitates to observe a similar 
trend for two conjugate measures as $r_c$ varies. Simultaneous escalation of both the ratios with increment of $r_c$ 
suggests delocalization in the system. Further, unlike Table~II, there is hardly any crossover among these $R_{\rvec}'$ and 
$R_{\pvec}'$ of $10l$ states (there were crossovers in $R_{\rvec}^{\alpha}$ in Table~II). More importantly, throughout the 
entire range of $r_c$, two ratios reduce with growth in $l$. This apparently indicates that states with greater number of 
nodes, experience the effects of confinement to a larger extent. Finally, in the limit of $r_c \rightarrow \infty$ these 
quantities for all $l$ states correspond to \emph{unity}, as expected. 

\begingroup           %%% table V
\squeezetable
\begin{table}
\caption{$S_{\rvec},~S_{\pvec}$ and $S_{t}$ for $2p,~3d$ states of CHA at various $r_c$. Reference values 
\cite{jiao17} for $2p$ and $3d$ states quoted here include angular contributions $S_{(\theta, \phi)}$, along with $S_r$ 
results given in the Supplementary Tables~VIII ($2p$) and XIV ($3d$) respectively. See text for details.}
\centering
\begin{ruledtabular}
\begin{tabular}{llll|llll}
$r_c$  &  $S_{\rvec} $ & $S_{\pvec} $  & $S_{t} $ &   $r_c$ &   $S_{\rvec} $  &    $S_{\pvec} $    & $S_{t}$   \\
\hline
\multicolumn{4}{c}{$2p$$^{\ddag,!}$}   \vline &      \multicolumn{4}{c}{$3d$$^{\ddag,\P}$}    \\
\hline
0.1  &  $-$6.3897304044 & 13.4182  & 7.0285       & 0.1  &  $-$6.3559834637   &  14.0037  & 7.6477          \\
0.2$^\S$  & $-$4.3126791264  & 11.3396  & 7.0270       & 0.2$^\dag$  &  $-$4.2772270811   &  11.9244   & 7.6472         \\
0.3  &  $-$3.0987163770  & 10.1240  & 7.0253       & 0.3  &  $-$3.0615253452   &   10.7082  & 7.6467         \\
0.5  &  $-$1.5712349891  &  8.5934  & 7.0222       & 0.5  &  $-$1.5304613573   &   9.1761   & 7.6457         \\
0.8  &  $-$0.1690563768  &  7.1867  & 7.0177       & 0.8  &  $-$0.1226356677   &  7.7668    & 7.6442         \\
1$^\S$  &  0.4949160183  &  6.5198  & 7.0147       & 1$^\dag$  &  0.5452929876   &  7.0980   & 7.6432         \\
5$^\S$  &  5.1596858896  &  1.83592 & 6.9956       & 7.5   &  6.5142648552   &   1.124506   & 7.63877         \\
15  &  7.2120252839  &  0.060324  & 7.272349     &  15  &  8.3843979910  &  $-$0.61531257  & 7.76908542      \\
25  &  7.2648183351  &  0.0423328 & 7.3071511    &  50$^\dag$  &  9.3456307493  &  $-$1.232729196 & 8.1129015533  \\
45  &  7.2648971157  &  0.04242079  & 7.3073179    & 100$^\dag$  &  9.3456341991  &  $-$1.232717246 & 8.112916953   \\
\end{tabular}
\end{ruledtabular}
\begin{tabbing}
$^\S$Literature results \cite{jiao17} of $(S_{\rvec}, S_{\pvec}, S_{t})$ at $r_c=0.2, 1, 5$ in $2p$ state are: 
($-$4.3126791236, 11.3396511782, 7.0269720545), \=   \\
(0.4949160211, 6.5198502152, 7.0147662362) and (5.1596858926, 1.8359293466, 6.9956152391) respectively.  \\
$^\dag$Literature results \cite{jiao17} of $(S_{\rvec}, S_{\pvec}, S_{t})$ at $r_c=0.2, 1, 50, 100$ in $3d$ state are: 
($-$4.2772270783, 11.9244780687,  6.8797080015), \>     \\ 
 (0.5452929905, 7.098078228, 7.6433712186), (9.3456307877, -1.2327291038, 8.112901684), \>     \\  
and (9.345634202, -1.2327172441, 8.112916958) respectively. \\
$^\ddag$Reference values are added with respective $S_{(\theta, \phi)}$ values (2.0990786249678, 2.0411250061339) of $2p$ 
and $3d$ states. \\
$^!$$S_\rvec,~S_\pvec,~S_t $ in FHA for $2p$ states ($|m|=0$) are: 7.264897118452,~0.042420799485,~7.307317917937. \\ 
$^\P$$S_\rvec,~S_\pvec,~S_t $ in FHA for $3d$ states ($|m|=0$) are: 9.345634202074,~$-$1.232717244109,~8.112916957965. 
\end{tabbing}
\end{table}
\endgroup
Table~III and S2 in SM now report our estimated values of $T_{\rvec}^{\alpha}, T_{\pvec}^{\beta}$ and $T_{t}^{(\alpha, \beta)}$ for $2p,3d$ and $4f,5g$ 
states of CHA successively at the same $r_c$'s of Table~I and S1. These are carefully selected so as to cover small, moderate and large cavity 
radius. Once again, there exists no reference work for comparison. In all these four states, starting from some ($-$)ve value, 
$T_{\rvec}^{\alpha}$'s advance monotonically with $r_c$ and finally reach the respective FHA behavior after some larger finite 
$r_c$. Like $R_{\rvec}^{\alpha}$ of Table~I and S1, $T_{\rvec}^{\alpha}$ is also negative up to $r_c \leq 0.5$. On the contrary, 
similar to $R_{\pvec}^{\beta}$ of Table~I and S1, as $r_c$ progresses, $T_{\pvec}^{\beta}$'s gradually decline from an initial result of 
$\approx \frac{1}{2}$, to large negative values (passing through a zero) at the end to merge with FHA result. Consequently, 
$T^{(\alpha, \beta)}$'s in these circular states grow (starting from a small negative) with advancement of $r_c$ and then attain a 
positive maximum and finally fall off to particular FHA value (large negative). As observed for $R$, $T$'s in both $r$, $p$ spaces 
follow the same order, \emph{viz.}, $T_{\rvec}^{\alpha}(5g)>T_{\rvec}^{\alpha}(4f)>T_{\rvec}^{\alpha}(3d)>T_{\rvec}^{\alpha}(2p)$ 
and  $T_{\pvec}^{\beta}(5g)>T_{\pvec}^{\beta}(4f)>T_{\pvec}^{\beta}(3d)>T_{\pvec}^{\beta}(2p)$ respectively. As usual, at 
$r_c \rightarrow \infty$ the order in $p$ space reverses to 
$T_{\pvec}^{\beta}(5g)<T_{\pvec}^{\beta}(4f)<T_{\pvec}^{\beta}(3d)<T_{\pvec}^{\beta}(2p)$. Then we proceed for 
$T_{\rvec}^{\alpha}, T_{\pvec}^{\beta}$ for $n=10$ states at same $r_c$ of Table~II. Similar to $R_{\rvec}^{\alpha}$, 
$T_{\rvec}^{\alpha}$'s remain ($-$)ve for all $l$ at $r_c < 1$ and ($+$)ve for $r_c > 1$. For each $l$, they increase
with $r_c$. Again analogous to $R_{\rvec}^{\alpha}$, at each $r_c$, $T_{\rvec}^{\alpha}$ first falls down to a minimum and 
finally grow with $l$. {\color{red} For first three $r_c$, these minima occur at $l=1$. For $r_c=10,~40$ these minima are 
found at $l=2$, whereas for $r_c=40$ and $100$ these minima appear at $l=3$ and $4$ successively.} On the other hand, $T_{\pvec}^{\beta}$ 
decrease with progress of $r_c$. Further, in all seven $r_c$, $T_{\pvec}^{\beta}$'s consistently decay with $l$, although the extent is 
very less till $r_c =1$ and assumes significance only after $r_c > 5$ or so. As there is a good resemblance of the behavior of CHA and 
FHA ratios in R\'enyi and Tsallis entropies in $r$ and $p$ spaces, one can expect and predict the qualitative nature of similar plots 
for $T$ for these states, and hence omitted here. 

\begingroup                         %%% table VI       
\squeezetable
\begin{table}
\caption{$S_{\rvec}$ and $S_{\pvec}$ for all $l$ states corresponding to $n=10$, of CHA at representative $r_c$ values, in top and 
bottom segments. For more details, consult text.}
\centering
\begin{ruledtabular}
\begin{tabular}{l|lllllll}
$l$  &  $r_c=0.1$ & $r_c=0.5$  & $r_c=1$  &   $r_c=10$   &  $r_c=40$  &   $r_c=80$  &    $r_c=100$      \\
\hline
\multicolumn{8}{c}{$S_{\rvec}$}    \\
\hline
0    &    $-$6.6336010412 &  $-$1.8032959712 &   0.2787117983  & 7.2442147026    & 11.5877476335   & 13.8154049673   & 14.5452382494   \\
1    &    $-$6.9827167662 &  $-$2.1537824035 &$-$0.0735403924  & 6.8533262967    & 11.1225960687   & 13.3570558294   & 14.0927363946   \\
2    &    $-$6.9592958229 &  $-$2.1308141271 &$-$0.0511509821  & 6.8629221900    & 11.0765559826   & 13.2799996428   & 14.0148698959   \\
3    &    $-$6.8974004176 &  $-$2.0691277981 &   0.0102684892  & 6.9184633212    & 11.0999960757   & 13.2630164065   & 13.9887469229   \\
4    &    $-$6.8231110824 &  $-$1.9949407785 &   0.0843246799  & 6.9895792871    & 11.1527172455   & 13.2805941590   & 13.9928101925   \\
5    &    $-$6.7421222230 &  $-$1.9139917010 &   0.1652222750  & 7.0691946490    & 11.2222138846   & 13.3235593289   & 14.0216304087   \\
6    &    $-$6.6566958653 &  $-$1.8285574196 &   0.2506653439  & 7.1545675577    & 11.3032904456   & 13.3866129318   & 14.0720358679   \\
7    &    $-$6.5693810669 &  $-$1.7411860014 &   0.3381068528  & 7.2431276172    & 11.3927769257   & 13.4664617170   & 14.1417536426   \\
8    &    $-$6.4873245243 &  $-$1.6590042222 &   0.4204449123  & 7.3282175171    & 11.4854342953   & 13.5598715121   & 14.2284568921   \\
9    &    $-$6.4437380685 &  $-$1.6151489690 &   0.4646366944  & 7.3785621302    & 11.5570097332   & 13.6533937166   & 14.3196849974   \\
\hline
\multicolumn{8}{c}{$S_{\pvec}$}    \\
\hline
0    &    18.225 &    13.399  & 11.324   & 4.710  & 0.711    & $-$1.206  & $-$1.8374  \\
1    &    17.685 &    12.858  & 10.7811  & 4.0149 & 0.4927   & $-$1.53   & $-$2.29   \\
2    &    17.512 &    12.684  & 10.6067  & 3.780  & 0.308    & $-$1.70   & $-$2.289  \\
3    &    17.367 &    12.539  & 10.461   & 3.605  & $-$0.032 & $-$1.78   & $-$2.412  \\
4    &    17.219 &    12.3919 & 10.313   & 3.440  & $-$0.28  & $-$1.872  & $-$2.557  \\
5    &    17.063 &    12.2355 & 10.156   & 3.270  & $-$0.581 & $-$2.09   & $-$2.6719  \\
6    &    16.889 &    12.061  & 11.982   & 3.087  & $-$0.875 & $-$2.427  & $-$2.906  \\
7    &    16.689 &    12.861  & 11.782   & 2.880  & $-$1.171 & $-$2.84   & $-$3.284  \\
8    &    16.447 &    12.618  & 11.539   & 2.632  & $-$1.482 & $-$3.31   & $-$3.792  \\
9    &    16.105 &    12.276  & 11.197   & 2.285  & $-$1.871 & $-$3.871  & $-$4.4326  \\
\end{tabular}
\end{ruledtabular}
\end{table}
\endgroup

\begin{figure}                         %%%Fig. 2, CHA
\begin{minipage}[c]{0.4\textwidth}\centering
\includegraphics[scale=0.75]{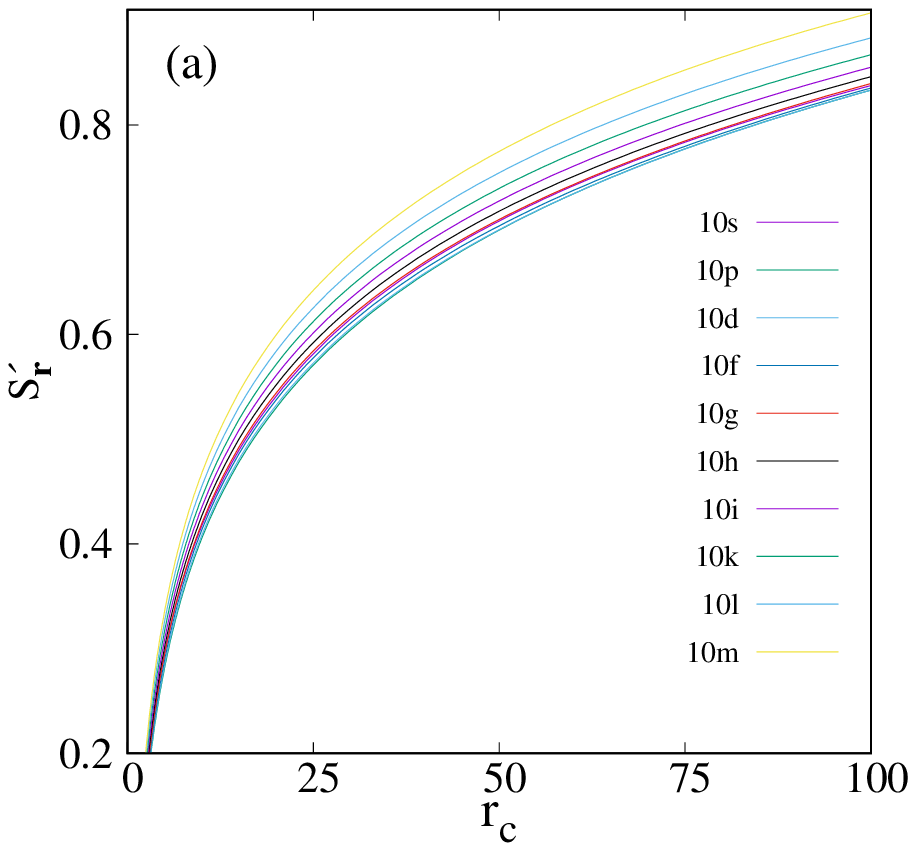}
\end{minipage}%
\hspace{0.1in}
\begin{minipage}[c]{0.5\textwidth}\centering
\includegraphics[scale=0.75]{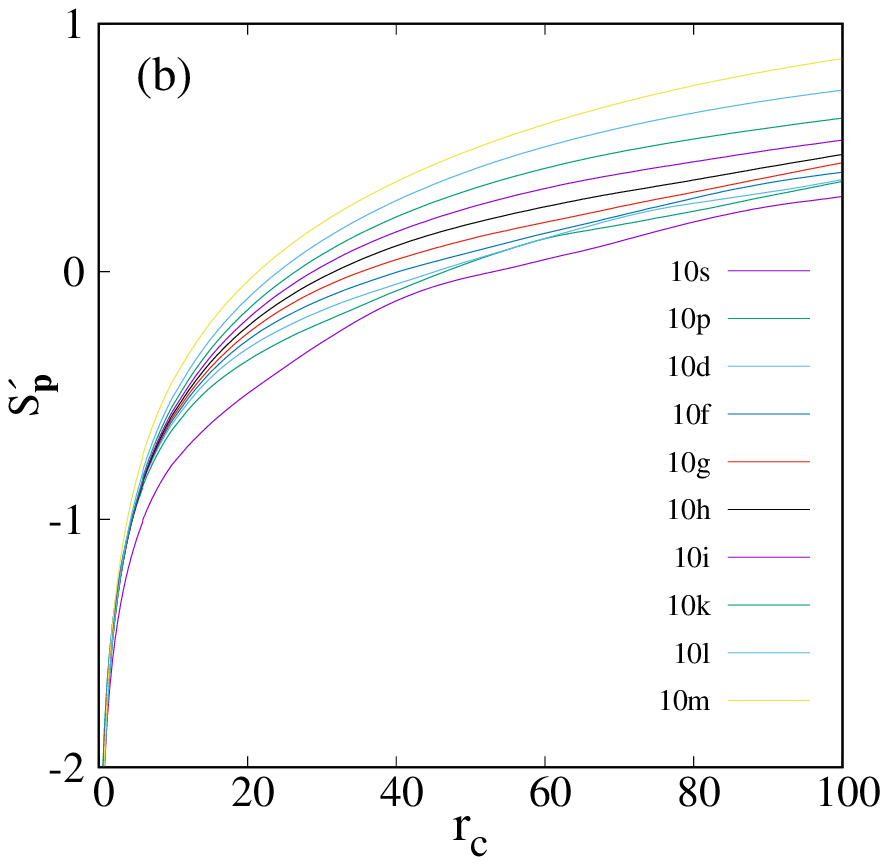}
\end{minipage}%
\caption{Variation of $S_{\rvec}', S_{\pvec}'$ of CHA with $r_c$, for $n=10$ states, in panels (a), (b). See text for detail.}
\end{figure}

Let us now shift our focus on Table~V and S3 in SM, where $S_{\rvec}, S_{\pvec}$ and $S_{t}$ of CHA are probed for four low-lying circular states 
corresponding to $l \! = \! 1-4$. The same set of $r_c$ of Table~I and S1 is adopted. This time, a handful of results are available 
in the literature for $2p$ (at $r_c=0.2,1,5$) and $3d$ ($r_c=0.2, 1, 50, 100$) states of CHA, which are duly quoted. Present 
results show good agreement 
with reference data in all occasions. $S_{\rvec}, S_{\pvec}, S_{t}$ portray analogous behavior to those of $R_{\rvec}^{\alpha}, 
R_{\pvec}^{\beta}$ and $R^{(\alpha, \beta)}$ respectively. Similar to $R_{\rvec}^{\alpha}$, $S_{\rvec}$ also take ($-$)ve 
values for all four states at $r_c < 1$ region and evolve continuously with growth of $r_{c}$ before reaching the FHA-limit 
at large $r_c$. On the contrary $S_{\pvec}$, like $R_{\pvec}^{\beta}$, shows a reverse nature of $S_{\rvec}$. $S_{\pvec}$ 
for $l \! = \! 1-4$ ($n \! = \! l+1$) states, starting from ($+$)ve numerical values, decline as $r_c$ develops before 
merging with FHA limit. Consequently, $S_{t}$, falls to reach a minimum and then elevates to attain FHA value. Furthermore, 
$S_{\rvec}$ imprints different pattern to $R_{\rvec}^{\alpha}$ but $S_{\pvec}$ delineates similar leaning to 
$R_{\pvec}^{\beta}$ in smaller $r_c$. The observed trend in $r, p$ spaces is slightly unusual: 
$S_{\rvec}(4f)>S_{\rvec}(5g)>S_{\rvec}(3d)>S_{\rvec}(2p)$ and 
$S_{\pvec}(5g)>S_{\pvec}(4f)>S_{\pvec}(3d)>S_{\pvec}(2p)$ respectively. As usual at $r_c \rightarrow \infty$ the order in 
$r, p$ space reorganize to $S_{\rvec}(5g)>S_{\rvec}(4f)>S_{\rvec}(3d)>S_{\rvec}(2p)$ and 
$S_{\pvec}(5g)<S_{\pvec}(4f)<S_{\pvec}(3d)<S_{\pvec}(2p)$. As a next step, Table~VI supplies $S_{\rvec}, S_{\pvec}$ for ten $l$
states corresponding to $n=10$ at same selected $r_c$ set discussed before. $S_{\rvec}$ remains ($-$)ve for all $l$ at first two
$r_c$'s. Interestingly, at $r_c=1$, $S_{\rvec}$ is ($-$)ve only for $l=1, 2$. In rest of the situation, $S_{\rvec}$ takes on 
($+$)ve values. $S_{\rvec}$, in all $r_c$'s at first diminish with $l$, attain some minima and then gains. {\color {red}
$S_{\rvec}$ reaches minimum at $l=1$ for first four $r_c$ values. At $r_c=40$ this mimimum shifts to $l=2$, whereas for $r_c=80$ and $100$ these 
minima arive at $l=3$.} But in $p$ space, $S_{\pvec}$ imitates
$R_{\pvec}^{\beta}$. Hence, like $R_{\pvec}^{\beta}$, here $S_{\pvec}$ collapses with rise of $l$. Here again the nature of 
$S_{\rvec}, S_{\pvec}$ variation for $10l$ states remains akin to respective $R_{\rvec}^{\alpha}, R_{\pvec}^{\beta}$ changes. Like 
other states reported here, $S_{\rvec}$ progresses and $S_{\pvec}$ reduces with growth of $r_c$.    

\begin{figure}                         %%%Fig. 3, CHA
\begin{minipage}[c]{0.32\textwidth}\centering
\includegraphics[scale=0.48]{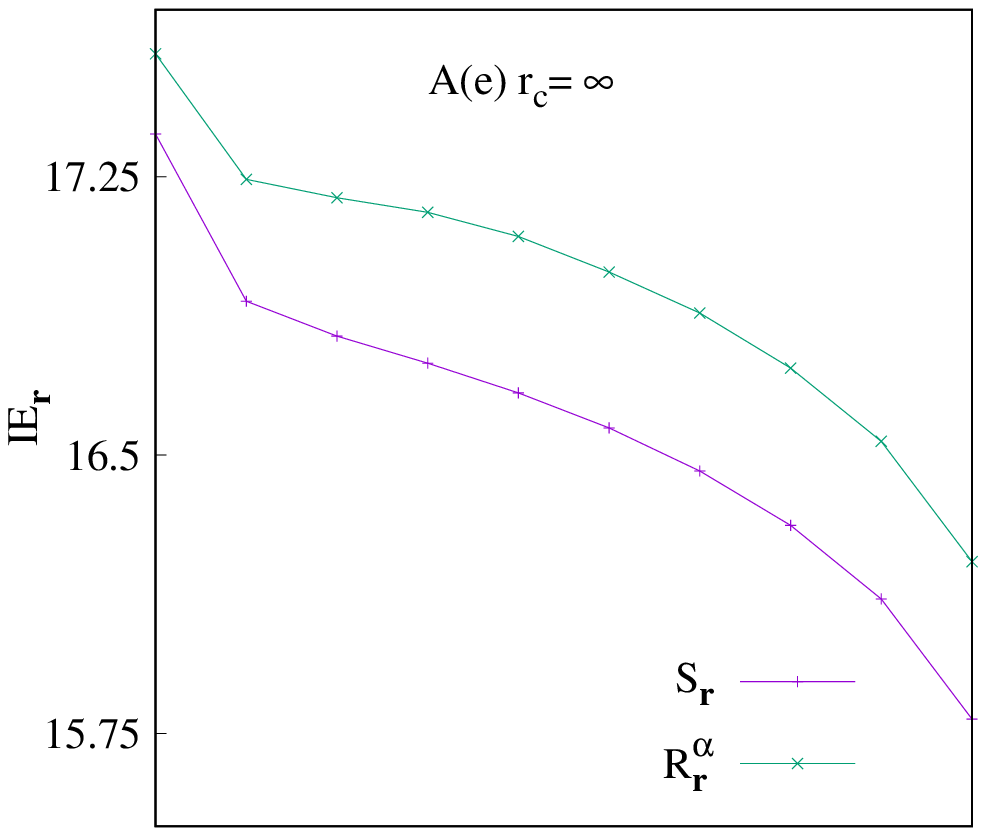}
\end{minipage}%
%\hspace{0.01in}
\begin{minipage}[c]{0.32\textwidth}\centering
\includegraphics[scale=0.48]{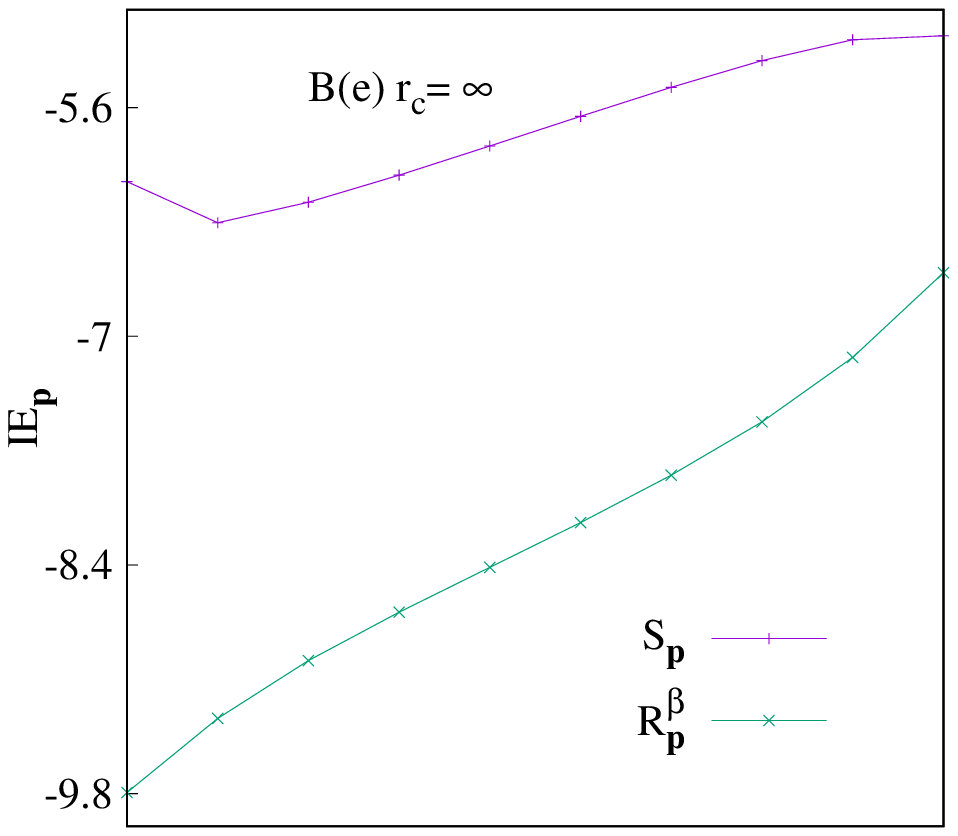}
\end{minipage}%
\hspace{0.01in}
\begin{minipage}[c]{0.32\textwidth}\centering
\includegraphics[scale=0.48]{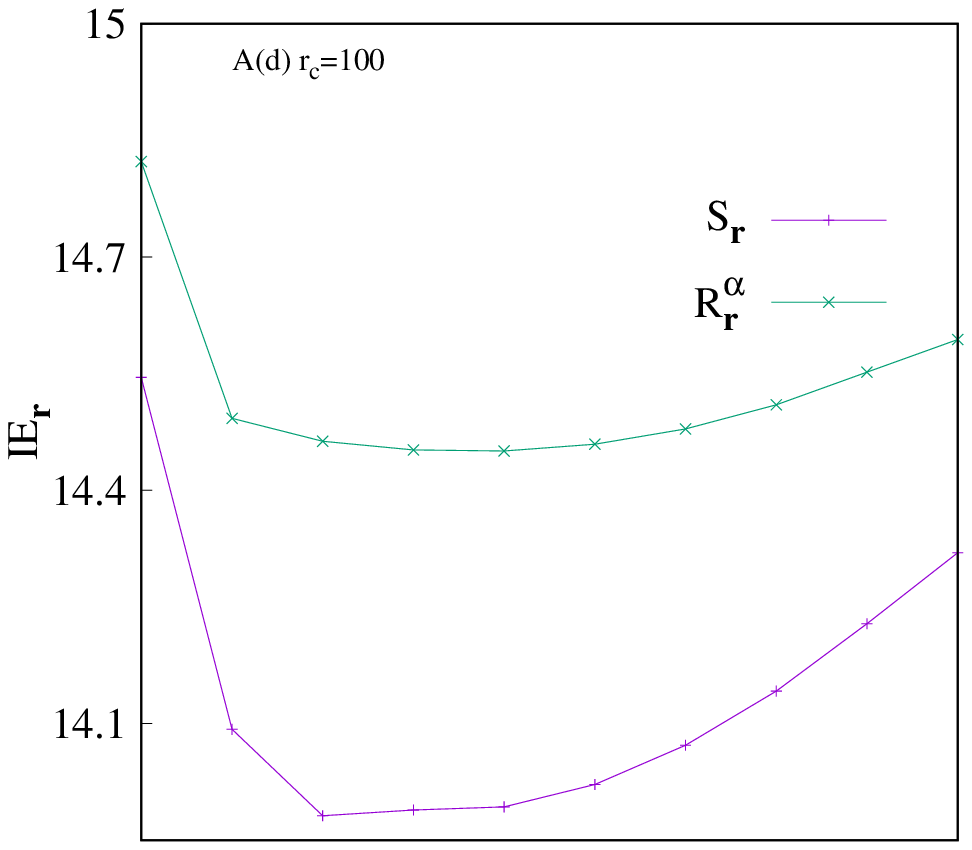}
\end{minipage}%
%\hspace{0.01in}
\begin{minipage}[c]{0.32\textwidth}\centering
\includegraphics[scale=0.48]{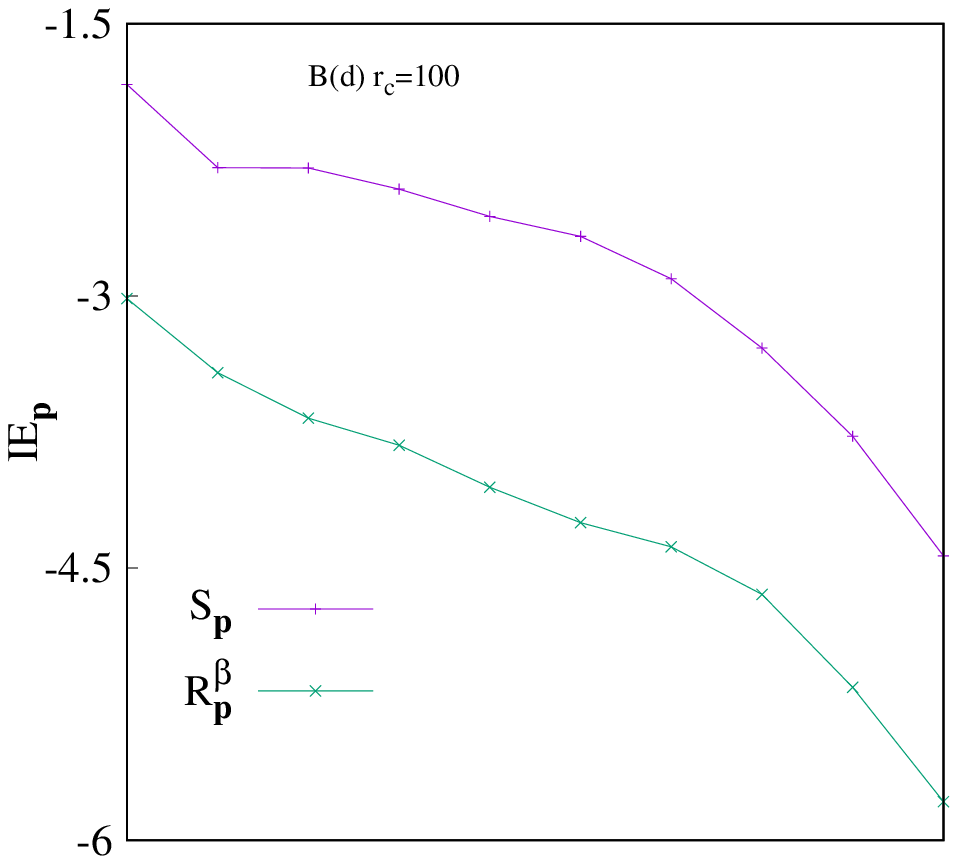}
\end{minipage}%
\hspace{0.01in}
\begin{minipage}[c]{0.32\textwidth}\centering
\includegraphics[scale=0.48]{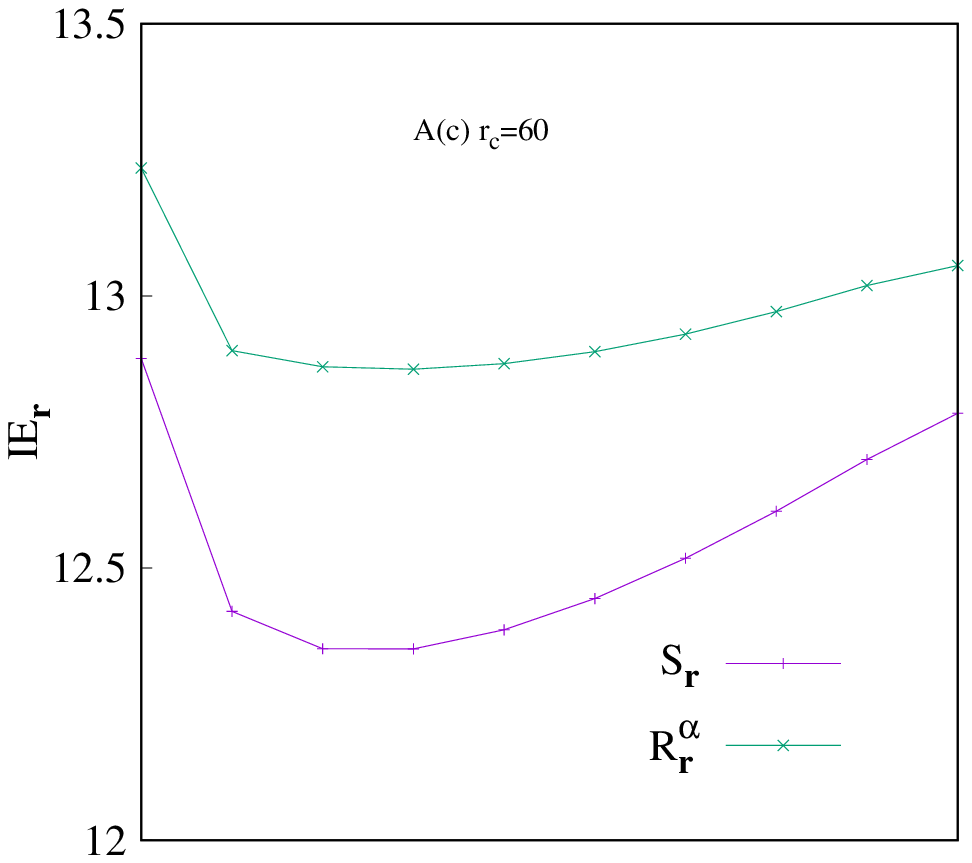}
\end{minipage}%
%\hspace{0.02in}
\begin{minipage}[c]{0.32\textwidth}\centering
\includegraphics[scale=0.48]{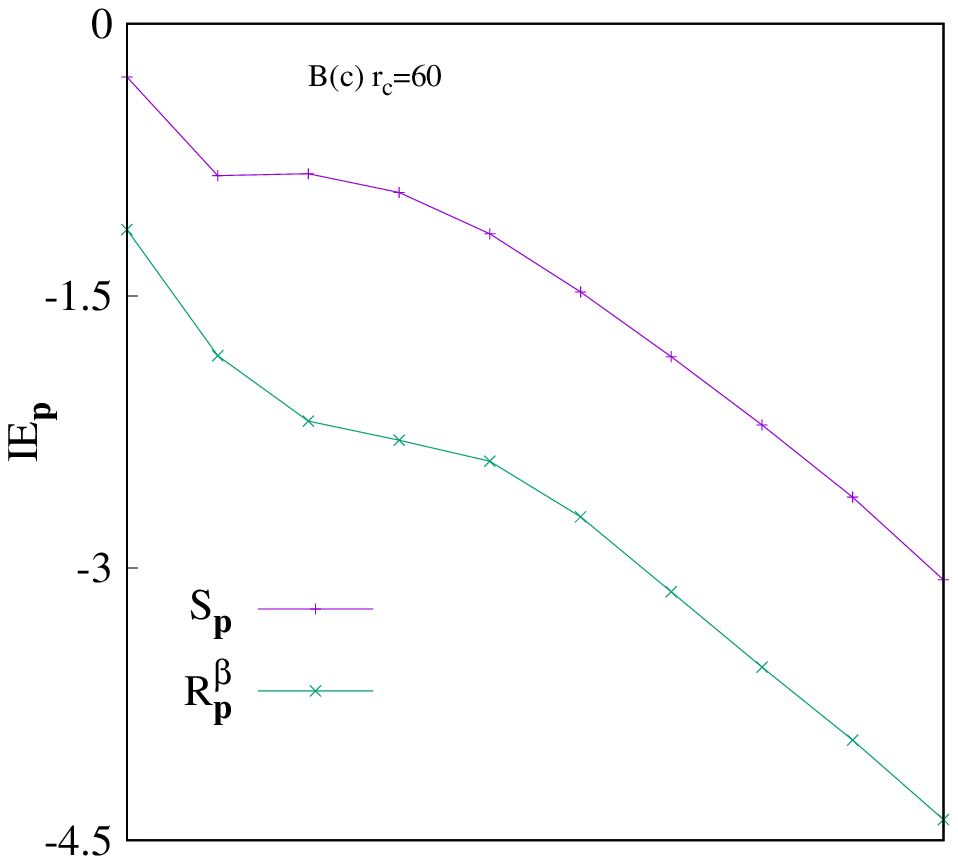}
\end{minipage}%
\hspace{0.01in}
\begin{minipage}[c]{0.32\textwidth}\centering
\includegraphics[scale=0.48]{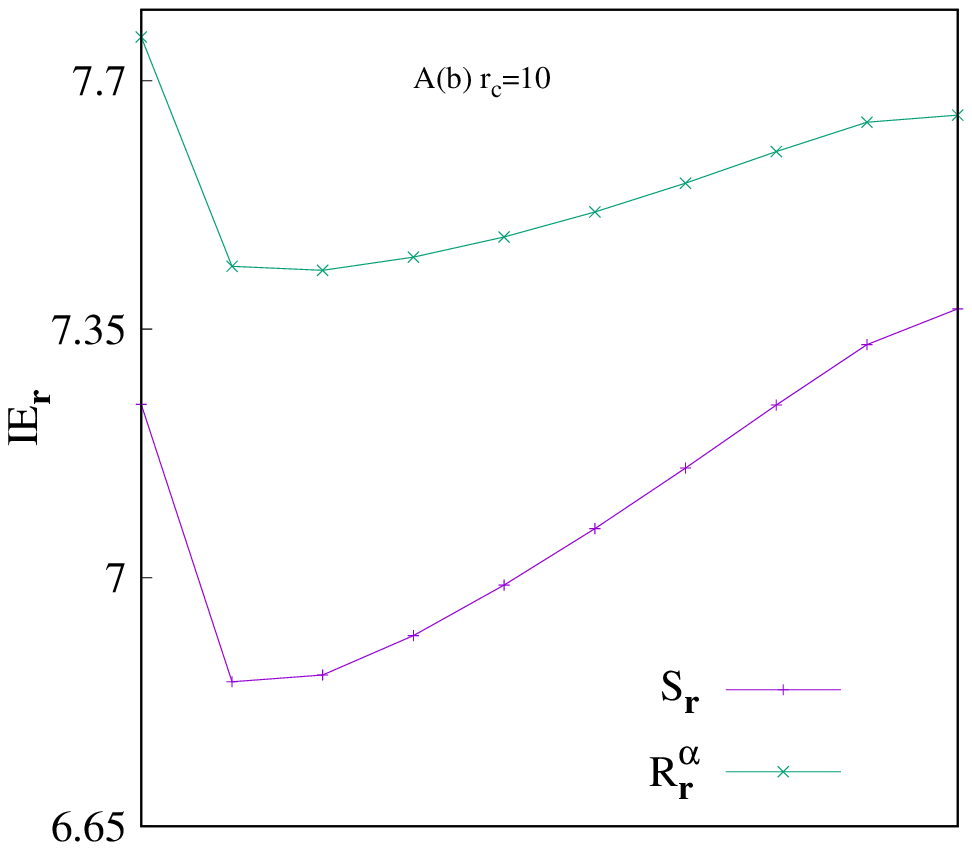}
\end{minipage}%
%\hspace{0.01in}
\begin{minipage}[c]{0.32\textwidth}\centering
\includegraphics[scale=0.48]{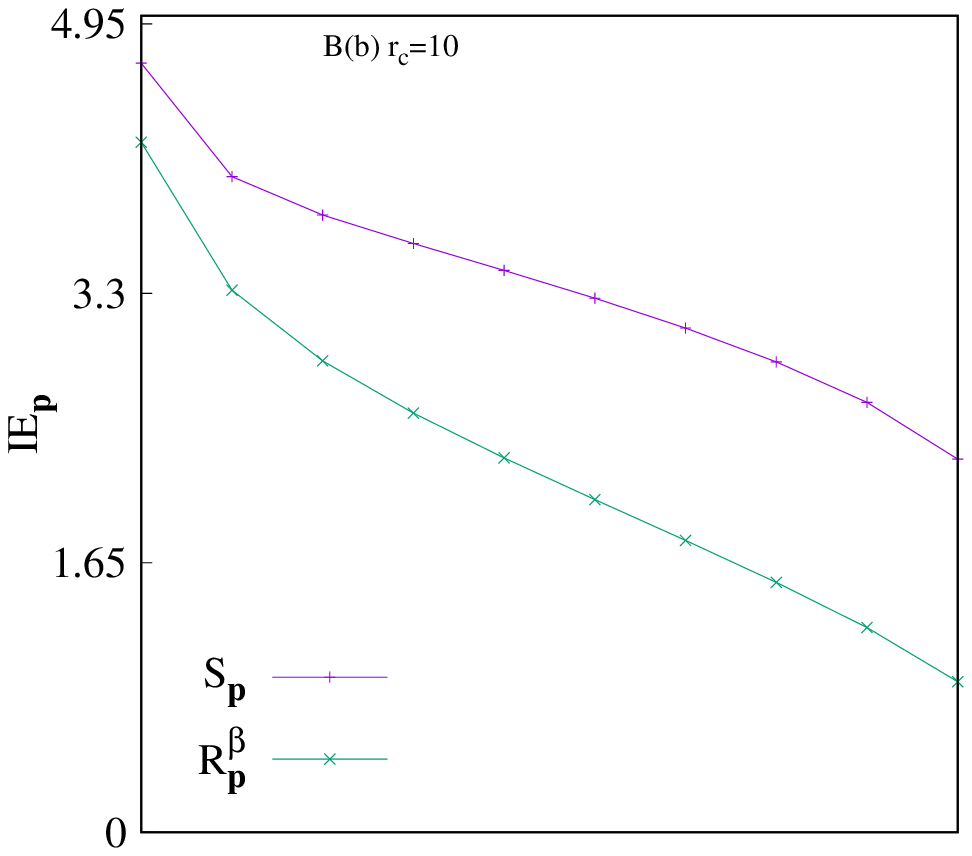}
\end{minipage}%0.0000017474
\hspace{0.01in}
\begin{minipage}[c]{0.32\textwidth}\centering
\includegraphics[scale=0.50]{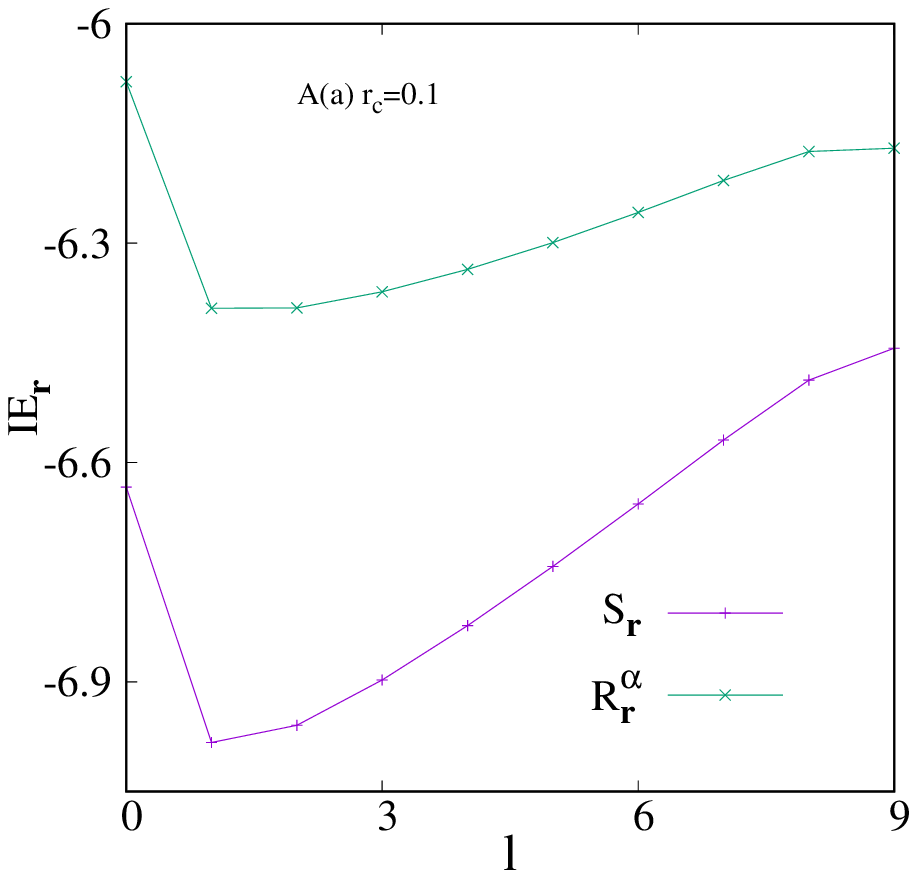}
\end{minipage}%
%\hspace{0.01in}
\begin{minipage}[c]{0.32\textwidth}\centering
\includegraphics[scale=0.50]{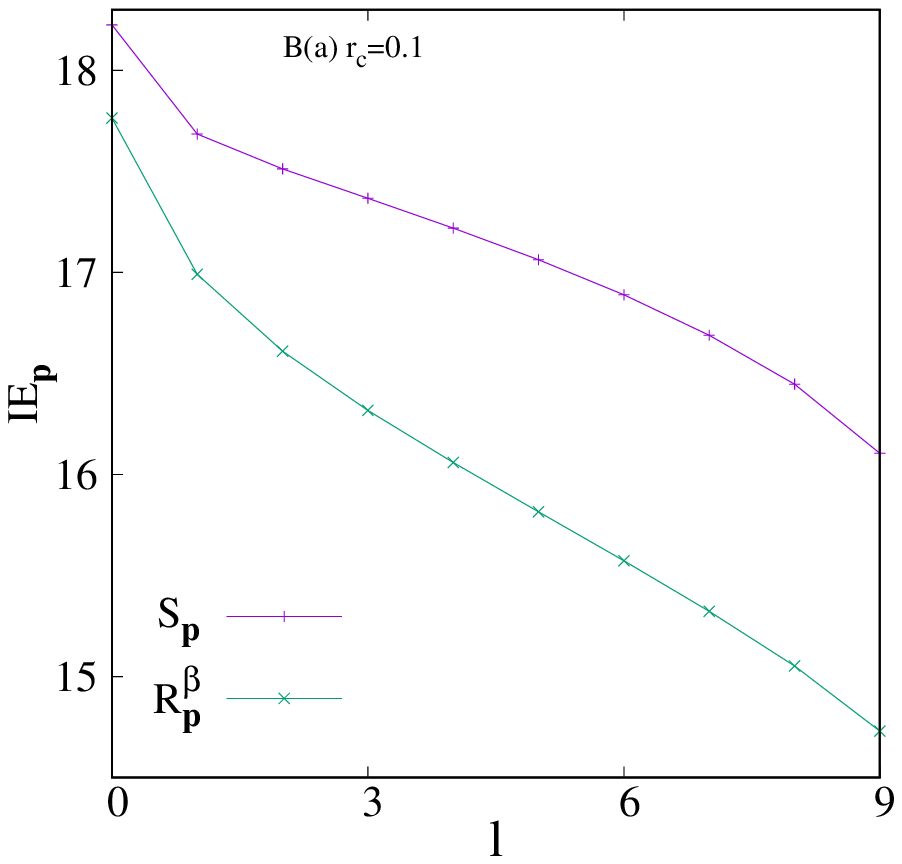}
\end{minipage}%
\caption{Variation of $R_{\rvec}^{\alpha}, S_{\rvec}$ ($IE_{\rvec}$) and $R_{\pvec}^{\beta}, S_{\pvec}$ ($IE_{\pvec}$) against 
$l$ for $n=10$ states of CHA in A and B columns. Five specific $r_c$'s (0.1, 10, 60, 100, $\infty$) are identified in 
parentheses (a), (b), (c), (d), (e) respectively. See text for detail.}
\end{figure}

Now we move on to Fig.~2 where following the strategy of Fig.~1, the relevant pair of unit-less ratios \emph{viz.}, 
$S_{\rvec}'=\left(\frac{S_{\rvec} (CHA)}{S_{\rvec}(FHA)}\right)$ and 
$S_{\pvec}'=\left(\frac{S_{\pvec}(CHA)}{S_{\pvec}(FHA)}\right)$ are displayed for all $l$ states having $n=10$. Drawing 
reference to Table~VI, one notices that, for any given $n, l$ quantum number, $S_{\rvec}$ rises and $S_{\pvec}$ falls 
continuously as $r_c$ proceeds to reach their corresponding maximum
and minimum limits at $r_c \rightarrow \infty$. Further, this limiting value in $S_{\pvec}$ relates to ($-$)ve sign (not 
obvious from Table~VI; but further extension of $r_c$ assures that). Thus both these ratios are bounded to their maximum 
values to unity. Hence, they both exhibit similar trend in behavior, i.e., grow up with $r_c$ signifying delocalization of
the system. In the entire range of $r_c$ the values of both $S_{\rvec}'$ and  $S_{\pvec}'$ abate with $l$, which suggests 
that, states with higher number of nodes, experience confinement to a greater extent. 

\begingroup            %%% table VII
\squeezetable
\begin{table}
\caption{$E_{\rvec}, E_{\pvec}, E_{t}$ for $2p, 3d$ states of CHA at various $r_c$. For more details, see text.}
\centering
\begin{ruledtabular}
\begin{tabular}{llll|llll}
$r_c$  &  $E_{\rvec} $ & $E_{\pvec} $  & $E_{t} $ &   $r_c$ &   $E_{\rvec} $  &    $E_{\pvec} $    & $E_{t}$   \\
\hline
\multicolumn{4}{c}{$2p^!$}    \vline &      \multicolumn{4}{c}{$3d^\P$}    \\
\hline
0.1  &  805.746259389130  & 0.00000227289 & 0.0018313726 & 0.1  &  851.986890492579  &  0.00000137674  & 0.0011729644   \\
0.2  & 101.040180954047   & 0.00001814520 & 0.0018333943 & 0.2  & 106.590831757296  &  0.00001100401  & 0.0011729266    \\
0.3  & 30.0353202004482  & 0.00006110958 & 0.0018354458 & 0.3  & 31.610240295272  &  0.00003710446  & 0.0011728809    \\
0.5 & 6.5311618804191  & 0.00028167391 & 0.0018396579 & 0.5   &  6.840055708558  &  0.00017145905  & 0.0011727895   \\
0.8  & 1.6113769393302 & 0.00114576730  & 0.001846263 & 0.8   & 1.674574222639  &  0.00070026869  & 0.0011726519   \\
1   &  0.8311099104230 & 0.00222698556 & 0.0018508698 & 1  & 0.859020100126  &  0.00136499761  & 0.0011725604   \\
5   &  0.0083268168966 & 0.23942780273 & 0.0019936715 & 7.5 &  0.002245898234  & 0.52169438759  & 0.0011716725  \\
15  & 0.0014334774547  & 1.90861683331 & 0.0027359592 & 15 &  0.000370433690  &  3.22168315792  & 0.00119342    \\
25  & 0.0013988503823  & 1.97640112599 & 0.0027646895 & 50 &  0.00001394232  &  35.52898842628  & 0.0004953565  \\ 
45  & 0.0013988227375  & 1.97576308974 & 0.0027637423 & 100 & 0.000172694164  & 7.10053419923  & 0.0012262208   \\
\end{tabular}
\end{ruledtabular}
\begin{tabbing}
$^!$$E_\rvec,~E_\pvec,~E_t $ in FHA for $2p$ states ($|m|=0$) are: 0.001398822737,~1.975763081024,~0.002763742330. \\ 
$^\P$$E_\rvec,~E_\pvec,~E_t $ in FHA for $3d$ states ($|m|=0$) are: 0.000172694164,~7.1005342704468,~0.0012262208417.
\end{tabbing}
\end{table}
\endgroup

Now Fig.~3 inscribes variation of $R_{\rvec}^{\alpha}, S_{\rvec}$ ($IE_{\rvec}$) and $R_{\pvec}^{\beta}, S_{\pvec}$ 
($IE_{\pvec}$) of $n=10$ states with change of $l$ in left and right columns labeled A, B. Here $IE$ stands for Information 
Entropy and in what follows this is loosely used to signify any or all of the measures discussed in this communication. These 
are given at five representative $r_c$'s namely, $0.1, 10, 60, 100, \infty$, identified by (a)-(e) in parentheses. At first 
four finite $r_c$'s, both $R_{\rvec}^{\alpha}$ and $S_{\rvec}$ fall off to reach certain minima and then improve with $l$ in 
panels A(a)-A(d). Positions of both these minima shift to right as $r_c$ is raised. Locations of these lowest points correspond 
to $l$ values in accordance with Tables~II and IV discussed before. However, in $r_c = \infty$ limit, for both measures in 
$r$ space, the minima disappear; rather there is a steady decline with $l$ which passes through a plateau region. The respective 
$p$-space quantities in first four panels B(a)-B(d) on the right side, decline gradually with $l$. However, in top fifth panel 
B(e), both of them are raised with increment of $l$. This study simply reveals distinctly separate behavior of a H atom 
from free to confined environment. In case of FHA, both spread and radial nodes of an wave-function reduce with $l$; hence 
$R_{\rvec}^{\alpha}$ as well as $S_{\rvec}$ diminish while $R_{\pvec}^{\beta}$ increases with uprise of $l$. Further, there appears a 
shallow minimum in $S_{p}$ for FHA. But in the confinement scenario, there exists an interplay between two mutually opposing factors: 
(i) radial confinement (favoring localization) and (ii) accumulation of radial nodes with reduction in $l$ (promoting delocalization). 
Hence, such minima appear in IE plots in panels A(a)-A(d) and B(a)-B(d). 

\begingroup            %%% table VIII
\squeezetable
\begin{table}
\caption{$E_{\rvec}$ and $E_{\pvec}$ for all $l$ states corresponding to $n=10$, of CHA at representative $r_c$ values, in top 
and bottom sections. For more details, consult text.}
\centering
\begin{ruledtabular}
\begin{tabular}{l|lllllll}
 \multicolumn{8}{c}{$E_{\rvec}$}    \\
\hline
$l$  &  $r_c=0.1$ & $r_c=0.5$  & $r_c=1$  &   $r_c=10$ & $r_c=40$  & $r_c=80$  &   $r_c=100$     \\
\hline
0    &    7747.631350106 &    62.389881644   & 7.861047634  &  0.008550277       & 0.0000975413   & 0.0000072401   & 0.0000029779   \\
1    &    6063.245277416 &    48.6353842949  & 6.0994482170 &  0.00641025319     & 0.0000996112   & 0.0000094350   & 0.0000040864    \\
2    &    4522.362533153 &    36.2403977184  & 4.5396157540 &  0.00470099482     & 0.0000773147   & 0.0000085650   & 0.0000039070    \\
3    &    3530.138368622 &    28.2761820347  & 3.5399931260 &  0.00363570042     & 0.0000603524   & 0.0000072797   & 0.0000034687    \\
4    &    2838.936113913 &    22.7331219908  & 2.8450195749 &  0.00290537538     & 0.0000480878   & 0.0000060612   & 0.0000029830    \\
5    &    2326.540075860 &    18.6259402411  & 2.3303725215 &  0.00236895765     & 0.0000389288   & 0.0000050118   & 0.0000025222    \\
6    &    1930.662444626 &    15.4535730094  & 1.9329918593 &  0.00195674165     & 0.0000318467   & 0.0000041310   & 0.0000021092    \\
7    &    1618.128828159 &    12.9493540013  & 1.6193470206 &  0.00163196755     & 0.0000262324   & 0.0000033940   & 0.0000017474   \\
8    &    1375.046723046 &    11.0013562943  & 1.3753245996 &  0.00137841885     & 0.0000217702   & 0.0000027783   & 0.0000014345   \\
9    &    1224.322475119 &    9.79175302714  & 1.2235277207 &  0.00121564677     & 0.0000186122   & 0.0000022894   & 0.0000011795    \\
\hline
\multicolumn{8}{c}{$E_{\pvec}$}    \\
\hline 
0    &    0.00000001732663 & 0.000002159204  & 0.000017186150  &    0.01275050    & 0.918902   & 6.52106   &  12.55735    \\
1    &    0.00000003453069 & 0.000004307362  & 0.00003435147   &    0.02945819    & 1.366699   & 10.40162  &  22.83099   \\
2    &    0.00000004594109 & 0.000005733162  & 0.00004575769   &    0.04151677    & 1.479949   & 13.08729  &  25.89834   \\
3    &    0.0000000570426  & 0.00000712045   & 0.00005685540   &    0.05313830    & 1.937268   & 15.702434 &  29.25821   \\
4    &    0.0000000693689  & 0.000008660817  & 0.00006917634   &    0.06584220    & 2.714584   & 16.70624  &  34.56930   \\
5    &    0.0000000840379  & 0.00001049398   & 0.00008383898   &    0.0807910     & 3.767910   & 18.52554  &  38.28028  \\
6    &    0.0000001024152  & 0.00001279075   & 0.0001022107    &    0.0993984     & 5.123417   & 24.45627  &  43.23403  \\
7    &    0.0000001266503  & 0.00001582009   & 0.0001264462    &    0.12388312    & 6.89702   & 37.097591 &  58.60068   \\
8    &    0.0000001609259  & 0.00002010565   & 0.0001607428    &    0.1585969     & 9.37541   & 59.3034   &  96.4456  \\
9    &    0.00000021939962 & 0.00002741951   & 0.00021930107   &    0.2181656     & 3.56075   & 99.301658 &  178.3531  \\
\end{tabular}
\end{ruledtabular}
\end{table}
\endgroup

Next we move on to discuss the last measure in this work, namely, $E$ for same four non-zero $l$ states as 
considered for $R, T, S$, in Table~VII and S4 in SM. Once again no literature results could be found to compare. Generally
speaking, behavior of $E$ is usually reverse to those of $R, T, S$ in Tables~I, III, V and S1-S3. Thus for all four states, $E_{\rvec}$'s
diminish, while $E_{\pvec}$ advance with surge of $r_c$. As $r_c$ approaches zero, $E_{\rvec}$ obeys the trend
$E_{\rvec}(5g)>E_{\rvec}(4f)>E_{\rvec}(3d)>E_{\rvec}(2p)$ which gets reversed to 
$E_{\rvec}(2p)>E_{\rvec}(3d)>E_{\rvec}(4f)>E_{\rvec}(5g)$ in opposite $r_c$ limit. Contrariwise, $E_{\pvec}$ shows exactly opposite 
trend from its $r$-space counterpart at both small and large $r_c$ regions. Finally, Table~VIII features $E_{\rvec}, E_{\pvec}$ 
for $n=10$ states in upper, lower portions. For all ten $l$, $E_{\rvec}$'s collapse with accrual of $r_c$. {\color{red} At first four $r_c$ values
$E_{\rvec}$ droops down with rise of $l$. But, in other three $r_c$ values there appear maxima at $l=1$.} In $p$ space, these patterns are 
completely reverse from $r$-space counterparts. Trends in $E_{\rvec}$, $E_{\pvec}$ with respect to $r_c$ complement the findings of $R, T, S$.     

\begin{figure}                         %%%Fig. 4, CHA
\begin{minipage}[c]{0.4\textwidth}\centering
\includegraphics[scale=0.75]{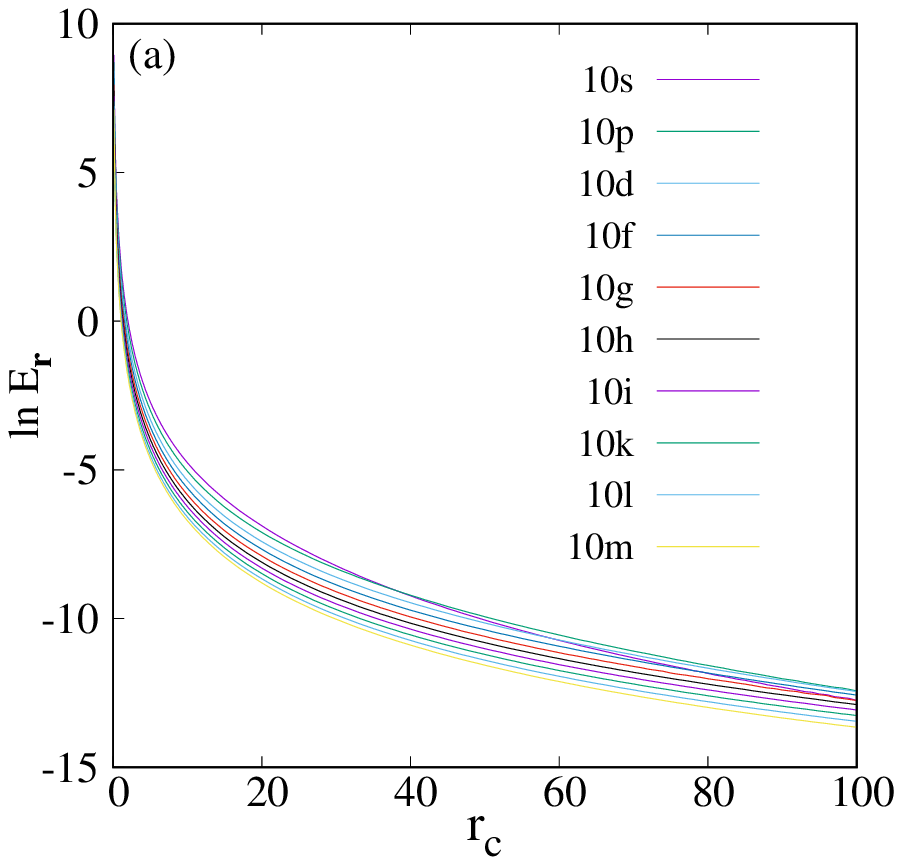}
\end{minipage}%
\hspace{0.1in}
\begin{minipage}[c]{0.5\textwidth}\centering
\includegraphics[scale=0.75]{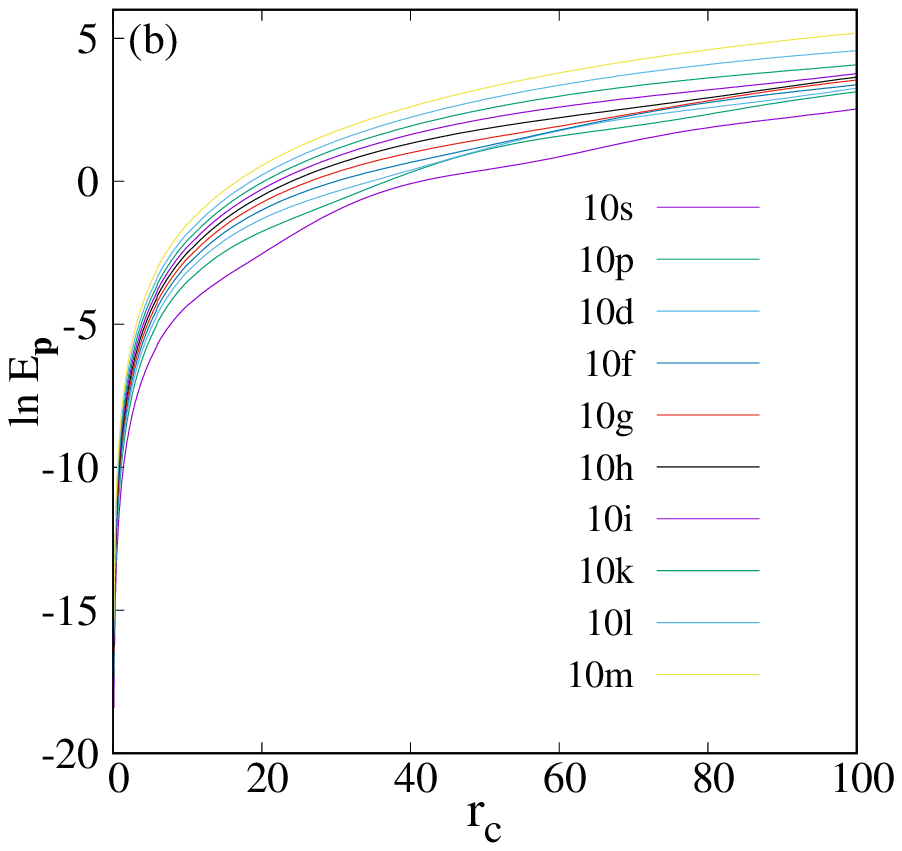}
\end{minipage}%
\caption{Logarithmic changes of $E_{\rvec}, E_{\pvec}$ of CHA with $r_c$, for $n=10$, in panels (a), (b). Consult text for detail.}
\end{figure}

Figure~4 now sketches the {\color{red} logarithmic} change of $E_{\rvec}$, $E_{\pvec}$ with $r_c$ for $n=10$ states in panels (a), (b). In contrast to $R$
and $S$, $E_{\rvec}$ lessens whereas $E_{\pvec}$ strengthens as $r_c$ grows, for any given $n,l$. Like all other measures, 
they both eventually merge to their FHA values. Convergence of $E_{\pvec}$ is not obvious from panel (b); however this can be 
verified upon extending $r_c$ to some sufficiently large value. As in case of Tables~VII and S4, this result again consolidates our 
previous discussion on $R, T, S$ that, relaxation in confinement facilitates  delocalization.  

Lastly in order to understand these measures with charge, Fig.~5 displays behavioral patterns of $R, S, E$ in conjugate 
$r,p$ spaces. In this occasion, we limit the discussion to ground state only, as it can be trivially extended for other states. 
These are followed at seven particular $Z$, \emph{viz.,} 1, 2, 3, 4, 5, 10, 15. Bottom and top rows characterizing
$r$- $p$-space properties are denoted by A, B, while $R, S, E$ are identified by parentheses (a), (b), (c) respectively. 
Note that  
$T$ behavior is similar in fashion to $R$; thus lead to common interpretation and hence not reported in this figure. It is seen 
from bottom three panels that, $R_{\rvec}^{\alpha}, S_{\rvec}$ advance, whereas, $E_{\rvec}$ falls off with an escalation in $Z$. 
Whereas from top row it is observed that $R_{\pvec}^{\beta}, S_{\pvec}$ decay while $E_{\pvec}$ intensifies with increment 
of $Z$. In B(c) segment, $Z=1, Z=2$ graphs also show similar behavior like other $Z$ values (not clear from panel B(c)). A careful 
scrutiny of these measures with respect to $Z$ suggests that, for any arbitrary $nl$ state, an increase in $Z$ promotes localization. 
Hence, the electron density gets tightened as one goes to heavier atoms. This strengthens the trend in results of various observed 
chemical phenomena like \emph{electronegativity}, \emph{ionisation potential}, \emph{hard-soft} interaction etc., in atomic 
systems. Table portraying the values of $R, T, S, E$ at various $Z$ are not given here. Because, such tables for $R, T, S, E$ can 
easily be constructed with the help of their definition in section II and data provided in section III.  

\begin{figure}                         %%%Fig. 5, CHA
\begin{minipage}[c]{0.3\textwidth}\centering
\includegraphics[scale=0.5]{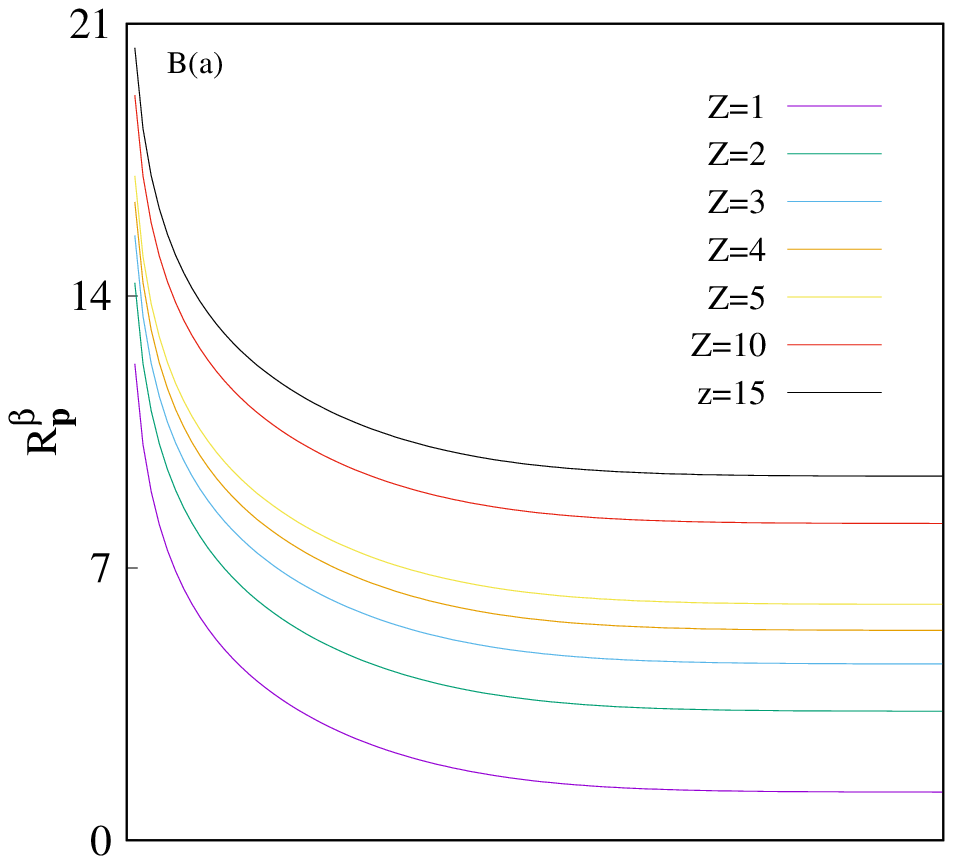}
\end{minipage}%
\hspace{0.02in}
\begin{minipage}[c]{0.3\textwidth}\centering
\includegraphics[scale=0.5]{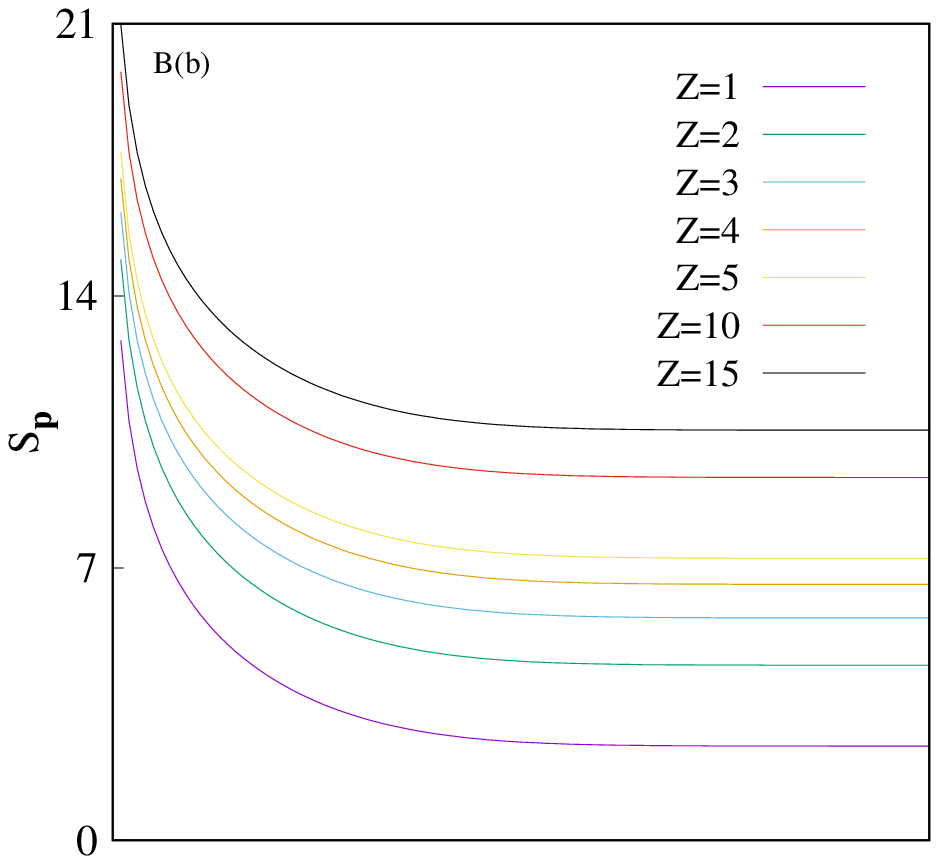}
\end{minipage}%
\hspace{0.02in}
\begin{minipage}[c]{0.3\textwidth}\centering
\includegraphics[scale=0.5]{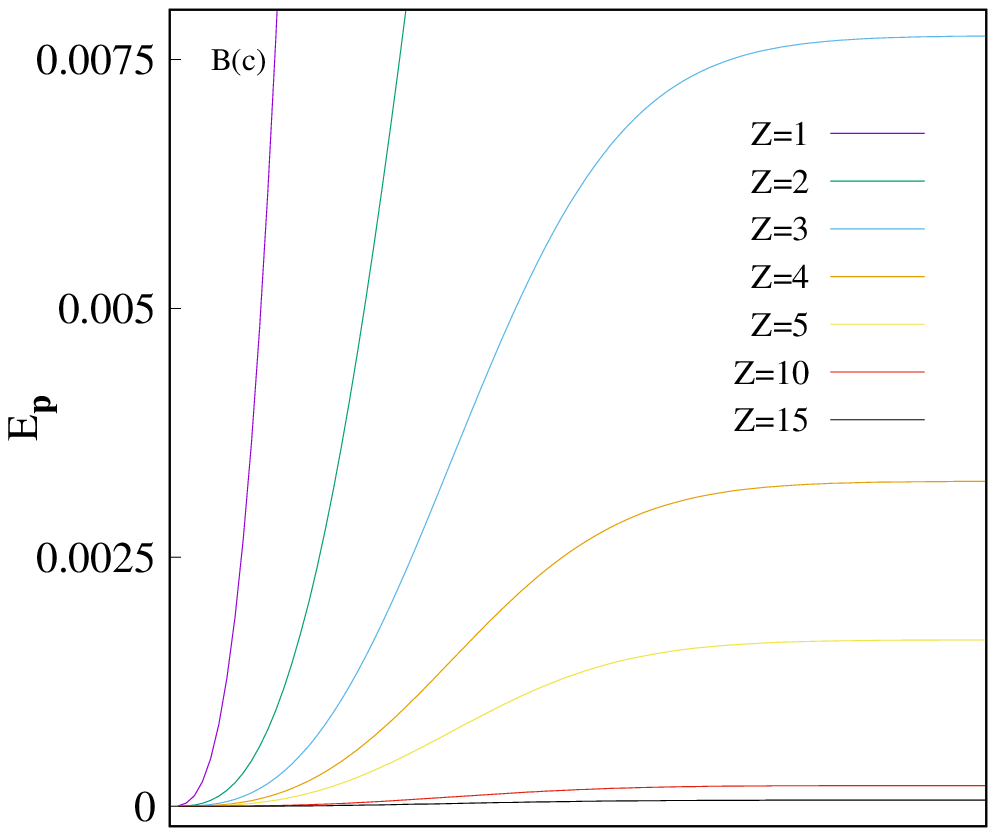}
\end{minipage}%
\hspace{0.02in}
\begin{minipage}[c]{0.3\textwidth}\centering
\includegraphics[scale=0.55]{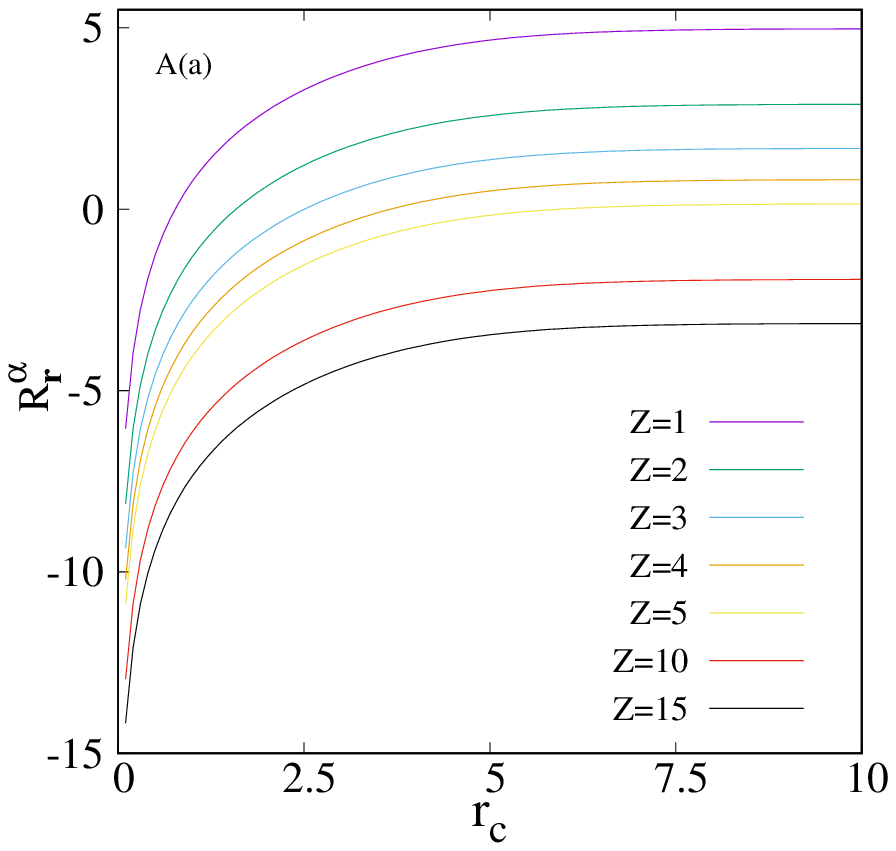}
\end{minipage}%
\hspace{0.02in}
\begin{minipage}[c]{0.3\textwidth}\centering
\includegraphics[scale=0.55]{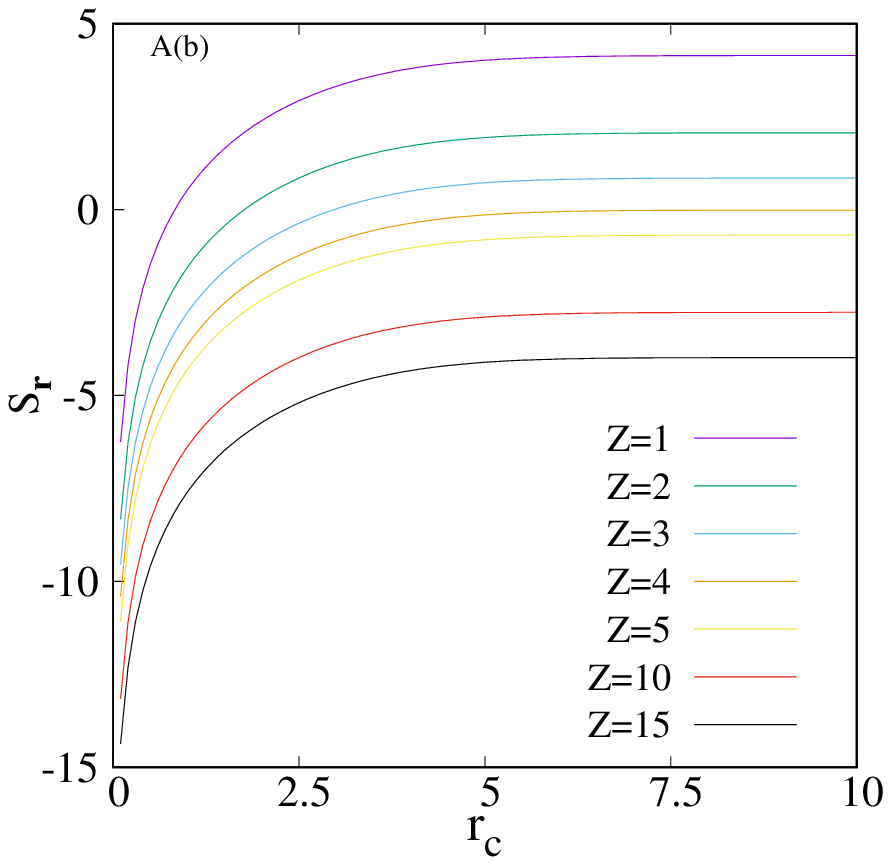}
\end{minipage}%
\hspace{0.02in}
\begin{minipage}[c]{0.3\textwidth}\centering
\includegraphics[scale=0.55]{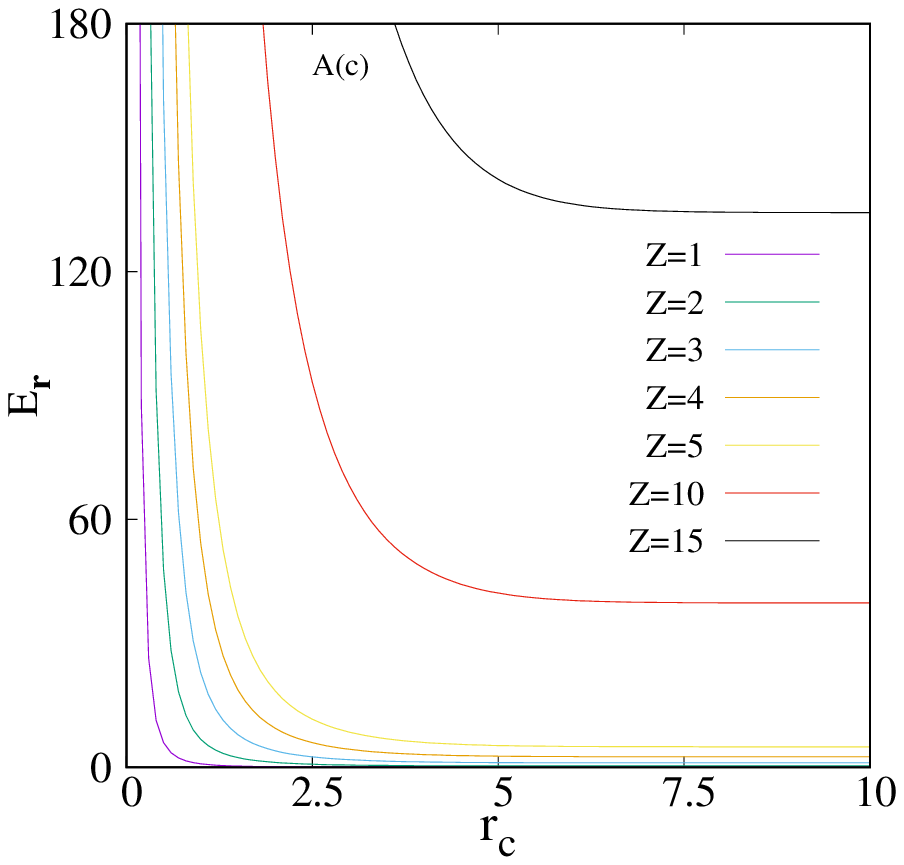}
\end{minipage}%
\caption{Nature of $R_{\rvec}^{\alpha}, S_{\rvec}, E_{\rvec}$ (A) and $R_{\pvec}^{\beta}, S_{\pvec}, E_{\pvec}$ (B) in $r$ (lower) 
and $p$ (upper) spaces, for seven selected $Z$. $R, S, E$ are labeled by parentheses (a), (b), (c) respectively. Ground-state 
results are given for $\alpha = \frac{3}{5}, \beta=3$. See text for details.} 
\end{figure}

\section{Future and Outlook}
Information-theoretic measures like $R,~T,~S,~E$ are explored for $l \neq 0$ states of CHA in both $r$, $p$ spaces. Accurate 
results for \emph{combined} measures (radial \emph{plus} angular) are provided for $2p, 3d, 4f, 5g$ and $n=10$ states of CHA, 
keeping $m$ fixed at zero. Except for very recent publication of $S_{\rvec}, S_{\pvec}$ in $2p, 3d$ states, all these quantities 
are reported 
for first time. It is found that at small $r_c$, with growth of $n$, $R_{\rvec}^{\alpha}, T_{\rvec}^{\alpha}$ build up while  
$E_{\rvec}$ deteriorate. Beside this, $R_{\rvec}^{\alpha}, T_{\rvec}^{\alpha}, S_{\rvec}$ pass through a minimum with addition of 
$l$. Further, an investigation on $n=10$ states has been made to get an idea about the high-lying states of CHA. It is realized
that, they 
may be exploited to analyze the spread as well as diffused nature of Rydberg-hydrogenic states. Additionally, scaling property has 
been utilized to ascertain the effect of \emph{atomic number} on IE of confined atomic systems. Actually, upgradation in $Z$ 
strengthens the $r$-space electron density of any arbitrary $nl$ state of CHA. An examination of these quantities in the realm 
of Rydberg states under various types of confined environment may be worthwhile to consider. A collateral inspection of 
many-electron atomic systems regarding IE and periodic properties would be highly desirable. 

\section{Acknowledgement}
Financial support from DST SERB, New Delhi, India (sanction order: EMR/2014/000838) is gratefully acknowledged. NM thanks DST SERB, 
New Delhi, India, for a National-post-doctoral fellowship (sanction order: PDF/2016/000014/CS).

\end{document}

%%%%%%%%%%%%%%%%%%%%%%%%%%%%%%% SUPPLYMENTARY MATERIAL %%%%%%%%%%%%%%%%%%%%%%%%%%%%%%%%%%%%%%%

\pagebreak
\widetext

\begin{center}
\textbf{\large Supplemental Materials: Information-entropic measures for non-zero $l$ states of confined hydrogen-like ions}
\end{center}

\setcounter{page}{1}
\setcounter{figure}{0}
\setcounter{table}{0}

\makeatletter
\renewcommand{\thefigure}{S\arabic{figure}}
\renewcommand{\thetable}{S\arabic{table}}

\vspace{1mm}

\begingroup            %%% table S1
\squeezetable
\begin{table}[h]
\caption{R\'enyi entropies, $R_{\rvec}^{\alpha},~R_{\pvec}^{\beta}$ and $R_{t}^{(\alpha,\beta)}$ for $4f,~5g$ states of 
CHA at selected $r_c$ values, with $\alpha= \frac{3}{5}$ and $\beta= 3$. See text for details.}
\centering
\begin{ruledtabular}
\begin{tabular}{llll|llll}
$r_c$  &  $R_{\rvec}^{\alpha}$ & $R_{\pvec}^{\beta}$ & $R_{t}^{(\alpha,\beta)}$ &   $r_c$ &   $R_{\rvec}^{\alpha} $  &    
$R_{\pvec}^{\beta} $    & $R_{t}^{(\alpha,\beta)}$   \\
\hline
\multicolumn{4}{c}{$4f^{!}$}   \vline &      \multicolumn{4}{c}{$5g^{\P}$}    \\
\hline
0.1  &  $-$6.09881032904  & 13.5563109       & 7.4575006     & 0.1  &  $-$6.09675536829 & 13.8239214     & 7.7271660      \\
0.2  &  $-$4.01952824354  & 11.4774071       & 7.4578789     & 0.2  &  $-$4.01733601542 & 11.7448104     & 7.7274744      \\
0.3  &  $-$2.80329426627  & 10.2615533       & 7.4582590     & 0.3  &  $-$2.80096355554 & 10.5287473     & 7.7277837     \\
0.5  &  $-$1.27114583083  &  8.7301707       & 7.4590249     & 0.5  &  $-$1.26853446452 &  8.9969400     & 7.7284055     \\
0.8  &   0.13835780349    &  7.3218298       & 7.4601876     & 0.8  &   0.14139955294   &  7.5879465     & 7.729346      \\
1    &   0.80744036310    &  6.6535319       & 7.4609723     & 1    &   0.81077543963   &  6.9192028     & 7.7299782     \\
7.5  &   6.83551829516    &  0.655708291     & 7.4912265862  & 7.5  &   6.85190763173   &  0.90113880    & 7.75304643    \\
15   &   8.87593577458    & $-$1.334419688   & 7.5415160866  &  25   &  10.42248270185   & $-$2.571994854 & 7.850487848  \\
75   &  11.39155328951    & $-$3.296164472   & 8.0953888175  &  95   &  12.54714379989   & $-$4.080092197 & 8.467051603  \\
100  &  11.39173125808    & $-$3.296184085   & 8.0955471731  & 100  &  12.54745436087   & $-$4.080121372 & 8.467332989   \\
\end{tabular}
\end{ruledtabular}
\begin{tabbing}
$^!$$R^{\alpha}_{\rvec},~R^{\beta}_{\pvec},~R_{t}^{(\alpha, \beta)}$ in FHA for $2p$ states ($|m|=0$) are: 11.391731416834988,~$-$3.2961840848475106,~8.095547331987477. \\ 
$^\P$$R^{\alpha}_{\rvec},~R^{\beta}_{\pvec},~R_{t}^{(\alpha, \beta)}$ in FHA for $3d$ states ($|m|=0$) are: 12.547737094562322,~$-$4.080150421410378,~8.467586673151944.
\end{tabbing}
\end{table}
\endgroup

\begingroup        %%% table S2    
\squeezetable
\begin{table}[h]
\caption{Tsallis entropies $T_{\rvec}^{\alpha},~T_{\pvec}^{\beta}$ and $T_{t}^{(\alpha,\beta)}$ for $4f,~5g$ states 
of CHA at various $r_c$, for $\alpha$, $\beta$ as $\frac{3}{5}$ and 3 respectively. For more details, see text.}
\centering
\begin{ruledtabular}
\begin{tabular}{llll|llll}
$r_c$  &  $T_{\rvec}^{\alpha}$ & $T_{\pvec}^{\beta}$  & $T_{t}^{(\alpha,\beta)} $ &   $r_c$ &   $T_{\rvec}^{\alpha} $  
       &    $T_{\pvec}^{\beta} $    & $T_{t}^{(\alpha,\beta)} $  \\
\hline
\multicolumn{4}{c}{$4f^{!}$}   \vline &      \multicolumn{4}{c}{$5g^{\P}$}    \\
\hline
0.1 & $-$2.2819941539  & 0.4999999999991 & $-$1.140997077   & 0.1  &  $-$2.281814882 & 0.4999999999995 & $-$1.140907441     \\
0.2 & $-$1.9991860306  & 0.4999999999463 & $-$0.9995930152  & 0.2  &  $-$1.998746678 & 0.4999999999685 & $-$0.999373339     \\
0.3 & $-$1.6853746581  & 0.4999999993892 & $-$0.8426873280  & 0.3  &  $-$1.684614841 & 0.4999999996420 & $-$0.842307420     \\
0.5 & $-$0.9964448564  & 0.4999999869370 & $-$0.4982224152  & 0.5  &  $-$0.994873502 & 0.4999999923382 & $-$0.497436744    \\
0.8 & 0.1422579964  & 0.4999997815714 & 0.0711289671     & 0.8  &  0.1454747876  & 0.4999998717186 &    0.0727373752      \\
1   & 0.9530810038  & 0.4999991686466 & 0.4765397096     & 1    &  0.9576905936  & 0.4999995113174 &    0.4788448288     \\
7.5 & 35.9938331267 & 0.3652809476268 & 13.1478614733    & 7.5  &  36.247017479  & 0.4175385840553 &   15.134528355      \\
15  & 84.5658772343 & $-$6.7116097699 & $-$567.573167850 &  25  &  159.12580417  & $-$85.199117451 & $-$13557.37807917   \\
75  & 235.6526959759 & $-$364.23890302 & $-$85833.8794779 &  95  & 375.59604544  & $-$1748.9158539 & $-$656885.8785564   \\
100 & 235.6696500573 & $-$364.25321055 & $-$85843.4266642 & 100  &  375.64301711 & $-$1749.0179353 & $-$657006.3742152   \\
\end{tabular}
\end{ruledtabular}
\begin{tabbing}
$^!$$T^{\alpha}_{\rvec},~T^{\beta}_{\pvec},~T_{t}^{(\alpha, \beta)}$ in FHA for $2p$ states ($|m|=0$) are: 235.6696405610288,$-$364.25321044734767,$-$85843.4231793272. \\ 
$^\P$$T^{\alpha}_{\rvec},~T^{\beta}_{\pvec},~T_{t}^{(\alpha, \beta)}$ in FHA for $3d$ states ($|m|=0$) are: 375.686384239036,$-$1749.1195832108906,$-$657120.411818189. 
\end{tabbing}
\end{table}
\endgroup

\begingroup           %%% table S3
\squeezetable
\begin{table}[h]
\caption{$S_{\rvec},~S_{\pvec}$ and $S_{t}$ for $4f,~5g$ states of CHA at various $r_c$ values. See text for details.}
\centering
\begin{ruledtabular}
\begin{tabular}{llll|llll}
$r_c$  &  $S_{\rvec} $ & $S_{\pvec} $  & $S_{t} $ &   $r_c$ &   $S_{\rvec} $  &    $S_{\pvec} $    & $S_{t}$   \\
\hline                             
\multicolumn{4}{c}{$4f^{!}$}   \vline &      \multicolumn{4}{c}{$5g^{\P}$}    \\
\hline
0.1 & $-$6.3455584854 & 14.4653  & 8.1198       & 0.1   &  $-$6.3503683331  &   14.8431  & 8.4928          \\
0.2 & $-$4.2663223577 & 12.3859  & 8.1196       & 0.2   &  $-$4.2709619968  &   12.7637  & 8.4928          \\
0.3 & $-$3.0501348575 & 11.1695  & 8.1194       & 0.3   &  $-$3.0546027165  &   11.5472  & 8.4926          \\
0.5 & $-$1.5180809495 & 9.6371   & 8.1191       & 0.5   &  $-$1.5222004720  &   10.0147  & 8.4925          \\
0.8 & $-$0.1087231165 & 8.2272   & 8.1185       & 0.8   &  $-$0.1123079582  &   8.6046   & 8.4923          \\
1   &  0.5602595052   & 7.5579   & 8.1181       & 1     &   0.5570394221   &   7.9351   & 8.4921           \\
7.5 &  6.5835558029   & 1.53173  & 8.11528      & 7.5   &  6.5968026117   &   1.89499  & 8.49179          \\
15  &  8.6122327603  & $-$0.460386  & 8.151846     &  25  & 10.1547840589  &   $-$1.581652  & 8.573132           \\
75  &  10.8608517807 & $-$2.14839726 & 8.71245452   &  95 & 12.0497760380  &   $-$2.86862362  & 9.18115241       \\
100 &  10.8608551968 & $-$2.14838651 & 8.71246868   & 100  & 12.0497927928  &  $-$2.86859762  & 9.18119517       \\
\end{tabular}
\end{ruledtabular}
\begin{tabbing}
$^!$$S_\rvec,~S_\pvec,~S_t $ in FHA for $4f$ states ($|m|=0$) are: 10.8608551994597,~$-$2.14838651125525,~8.712468688204. \\ 
$^\P$$S_\rvec,~S_\pvec,~S_t $ in FHA for $5g$ states ($|m|=0$) are: 12.0498008635736,~$-$2.8685790066675,~9.1812218569060.
\end{tabbing}
\end{table}

\begingroup            %%% table S4
\squeezetable
\begin{table}[h]
\caption{$E_{\rvec}, E_{\pvec}, E_{t}$ for $4f, 5g$ states of CHA at various $r_c$. For more details, see text.}
\centering
\begin{ruledtabular}
\begin{tabular}{llll|llll}
$r_c$  &  $E_{\rvec} $ & $E_{\pvec} $  & $E_{t} $ &   $r_c$ &   $E_{\rvec} $  &    $E_{\pvec} $    & $E_{t}$   \\
\hline     
\multicolumn{4}{c}{$4f^{!}$}   \vline &      \multicolumn{4}{c}{$5g^{\P}$}    \\
\hline
0.1  & 902.2306730117579  &  0.00000092547  & 0.0008349874 & 0.1  & 955.29622436210  &  0.00000066689 & 0.0006370775   \\
0.2  & 112.8081519710870  &  0.00000740021  & 0.000834804  & 0.2  &  119.41827435154 &  0.00000533361 & 0.0006369305   \\
0.3  & 33.4334274966356   &  0.00002496358  & 0.000834618  & 0.3  & 35.38507934013   &  0.00001799573 & 0.0006367803   \\
0.5  & 7.2254852493440   &  0.00011545852  & 0.0008342438 & 0.5  & 7.64401596722   & 0.00008326490 & 0.0006364782     \\
0.8 &  1.7654919808300   &  0.00047220577  & 0.0008336755 & 0.8  & 1.86653668200   & 0.00034074934 & 0.0006360211    \\
1   &  0.9044440955343  &  0.00092133060  & 0.000833292  & 1  & 0.95578172600  &  0.00066512447 & 0.0006357138   \\
7.5 &  0.0022027597662  & 0.37170465631  & 0.0008187761 & 7.5  &  0.00228068479  & 0.27382597673 & 0.0006245107  \\
15  &  0.0002938424484  &  2.71341384867 & 0.0007973162 & 25  &  0.00006598823  &  8.79992837948 & 0.0005806917  \\
75  & 0.0000384084606  & 17.97191418384  & 0.0006902736 & 95  &  0.00001189257  &  37.45698566645 & 0.0004454598  \\
100 &  0.000038408429927  & 17.97194614910  & 0.0006902742 & 100  &  0.00001189251  &  37.45679452927 & 0.0004454553  \\
\end{tabular}
\end{ruledtabular}
\begin{tabbing}
$^!$$E_\rvec,~E_\pvec,~E_t $ in FHA for $4f$ states ($|m|=0$) are: 0.000038408430432,~17.97194626972933,~0.00069027424804. \\ 
$^\P$$E_\rvec,~E_\pvec,~E_t $ in FHA for $5g$ states ($|m|=0$) are: 0.0000118924912048,~37.456994650567,~0.000445456979443.
\end{tabbing}
\end{table}
\endgroup